\documentclass[10pt,a4paper,twoside]{article}
\pdfoutput=1
\usepackage{a4wide}

\usepackage[utf8]{inputenc} 
\usepackage[TS1,T1]{fontenc}

\usepackage{amssymb, amsmath,subeqnarray}
\usepackage{mathrsfs}
\usepackage{bbold}
\DeclareFontFamily{U}{bbold}{}
\DeclareFontShape{U}{bbold}{m}{n}{<-5.5> bbold5 <5.5-7.5> bbold7 <7.5-> bbold10}{}

\makeatletter

\@addtoreset{equation}{section}
\makeatother

\usepackage{graphicx}

\usepackage[french,english]{babel}

\newcommand{\be}{\begin{equation}}
\newcommand{\ee}{\end{equation}}
\newcommand{\bea}{\begin{eqnarray}}
\newcommand{\eea}{\end{eqnarray}}

\newcommand{\Aev}{\mathbb{A}}

\newcommand{\C}{{\cal C}}

\newcommand{\D}{\mathbb{D}}

\newcommand{\Esp}[1]{\langle#1\rangle}

\newcommand{\Espst}[1]{\langle#1\rangle_\text{\mdseries st}}
\newcommand{\Extent}{\xi}

\newcommand{\intp}{\hat{\int}}

\newcommand{\Id}{\mathbb{I}}

\newcommand{\K}{\mathbb{K}}

\newcommand{\longc}{M}
\newcommand{\M}{\mathbb{M}}
\newcommand{\N}{\mathbb{N}}

\newcommand{\Oobs}{{\mathcal O}}

\newcommand{\Pdt}{\mathbb{P}}
\newcommand{\Pit}{\mu}
\newcommand{\Pin}{\nu}
\newcommand{\Ppdt}{\check{\mathbb{P}}}
\newcommand{\Prob}{P}

\newcommand{\Probst}{P_\text{\mdseries st}}

\newcommand{\Q}{\mathbb{Q}}

\newcommand{\SG}{S^{\scriptscriptstyle SG}}

\newcommand{\Trans}{\mathbb{W}}

\newcommand{\Uev}{\mathbb{U}}

\newcommand{\jQ}{j^{\Q}}
\newcommand{\XQ}{X^{\Q}}
\newcommand{\jS}{j^{\mathbb{S}}}
\newcommand{\XS}{X^{\mathbb{S}}}

\title{\textbf{ Affinity and Fluctuations in a Mesoscopic Noria} }
\author{
  M. Bauer\\
  Institut de Physique Théorique de Saclay\footnote{CEA/DSM/IPhT, URA2306 du
    CNRS}, CEA Saclay \\ F-91191 Gif-sur-Yvette Cedex, France
  \\ \vspace{3mm}\\
  F. Cornu\\ Laboratoire de Physique Théorique,  UMR 8627 du CNRS\\
  Université Paris-Sud, Bât. 210 \\ F-91405 Orsay, France }

\date{\today}

\begin{document}
\maketitle

\begin{abstract}
  We exhibit the invariance of cycle affinities in finite state Markov processes
  under various natural probabilistic constructions, for instance under
  conditioning and under a new combinatorial construction that we call
  ``drag and drop''. We show that cycle affinities have a natural probabilistic
  meaning related to first passage non-equilibrium fluctuation relations that we
  establish.

\vspace{.5cm}


\vspace{.5cm}
 
{\it Corresponding author ~:} \\
BAUER Michel\\
Fax :  01 69 08 81 20 \\
Email : michel.bauer@cea.fr
\end{abstract}

\section*{Introduction}

Affinity, a term ultimately borrowed from alchemy where it described a ``natural
attraction'' among certain elements, and later seen in early chemistry as the
``force'' that causes chemical reactions, has by now a precise thermodynamical
definition, related to the rate of irreversible variation of entropy when a
chemical reaction progresses \cite{Prigogine1968}. By analogy, affinity can be
defined in a much more general context for certain random processes describing
the time evolution of a probability measure, which offers striking analogies with
the time evolution of concentrations of reactants: the variation of the entropy
of the probability measure as time goes by often splits in a natural way and
allows to single out irreversible entropy variations, hence affinities
\cite{Schnakenberg1976}.

\vspace{.2cm}

The purpose of this study is twofold. 

-- The first is to study important invariances of affinities: affinities for
single ``reactions'' are in general not invariant under natural probabilistic
constructions, but affinities for cycles of reactions are much more robust
objects. We study this invariance for two important examples. The first, related
to a combinatorial construction which we call ``drag and drop'' is tailored for
this study and is explained in detail. The second, conditioning, relies on
well-known universal principles. Both of them allow, starting from a finite
state Markov process, to define natural processes
satisfying certain constraints, while preserving cycle affinities.

-- These observations allow us to obtain a probabilistic interpretation of cycle
affinities by concentrating on a single cycle even if the initial pattern of
reactions is more general. This interpretation is our second aim, and we show
that it leads to relations similar (or better dual) to standard
out-of-equilibrium fluctuation relations \cite{AndrieuxGaspard2007JStatPhys}:
the efficiency of a cycle (i.e. the preferred direction in which it is
traversed, and the speed at which this traversal is made) is quantified by
studying the winding number, and the cycle affinity is shown to be the crucial
observable that relates the probability distribution of the time it takes to
observe a winding number or its opposite. 

\vspace{.2cm}

The origin of the term ``dual'' is that instead of looking at fluctuations of an
observable at a given time, we look at the fluctuations of the time it takes to
reach a certain value of an observable. The dual picture has some relevance:
experimentally, looking at the fluctuations of the time it takes to observe
something is in fact quite common.  For instance certain experimental setups are
better suited to measure the time it takes for a particle to reach a certain
displacement than to measure the displacement at a fixed time.

\vspace{.2cm}

This article is organized as follows:
 
-- Section \ref{sec:ati} contains mostly standard background material. It starts
by motivating and introducing the basic notions related to affinities in the
context of Markov processes: cumulative processes, exchange quantities and
processes, exchange currents. On top of their physical importance, those objects
also have deep roots in probability theory.  Restricting our attention to
micro-reversible Markov processes (for which an allowed transition from
configuration $\C$ to configuration $\C'$ goes together with an allowed
transition from configuration $\C'$ to configuration $\C$) we recall how
affinity is related to a canonical (because it involves only that data of the
Markov process itself) exchange process via entropy and entropy variations: we
reproduce the standard computations that motivate this interpretation and add a
seemingly unknown remark on the positivity of the affinity current itself
(before average) in the stationary state (see \eqref{eq:poscur}). From the
affinity, which is defined for each reaction, i.e.  each unoriented edge of the
graph associated to the Markov process, we review the construction of the cycle
affinity and its invariances, and define a convenient technical tool which we
call the affinity class.

-- Section \ref{sec:acpc} shows that many natural probabilistic operations on
Markov processes preserve cycle affinities and the affinity class. We introduce
a combinatorial construction which we call ``drag and drop''. It associates to
every walk on a graph a walk on a given subgraph.  Then we explain what kind of
process is induced by ``drag and drop'' applied to samples of a Markov
process. It turns out that the continuous time Markov property is lost, but a
generalized renewal property survives, which is enough for all our purposes.  We
also review briefly the more familiar construction of conditioning. ``Drag and
drop'' and conditioning amount, albeit in different ways, to discard certain
reactions (i.e. transitions) occurring in the original Markov process. We then give
a list of probabilistic constructions, including ``drag and drop'' and
conditioning, that preserve the affinity class. We conclude with some remarks on
the observability of conditioning and ``drag and drop'' in real systems.

-- Section \ref{sec:pican} concentrates on the important case when the
transitions in the system define a single cycle. The relevance of this simple
case is enhanced by our previous observations of the good behavior of cycle
affinities when edges are discarded either by conditioning or by ``drag and
drop'': what we say for a single cycle remains true for a cycle embedded in a
more general pattern of reactions. In an out-of-equilibrium situation, a
current, which deserves the name winding current, will flow through the cycle.
This current is associated to an exchange process which is indeed the winding
number. After an algebraic preliminary, we give a heuristic argument to show
that if the affinity $A$ is $\geq 0$, then the probability that the winding
number will ever reach $-1$ is $e^{-A}$ while the probability to reach any
positive winding number is $1$. Then we refine the heuristic argument to prove
this result, and a stronger one related to first passage times and which is a
simple dual fluctuation relation involving the affinity again (see
\eqref{eq:flucrel}, which can be considered as the main equation in this work).
We conclude with a formula \eqref{eq:afpt} for the mean of the first passage
time at winding number $1$, a quantitative measure of efficiency. As already
said, instead of looking at the distribution of an observable at a given time,
we look at the distribution of the time it takes to reach a certain value of an
observable. The two approaches are related, and this allows us to check the
formula \eqref{eq:afpt} for the mean of the first passage time against the
mean of the winding at large times. More generally, though we have not
tried to prove it, there must exist relationships between the fluctuation
relations derived here and those obtained in
\cite{AndrieuxGaspard2007JStatPhys}.

\vspace{.2cm}

A few basic notions, included for completeness and to fix notations, are
gathered in two appendices. \textit{The reader is advised to look in the
  appendices whenever in trouble with a result used in the main text, or with
  terminology}.

-- Appendix \ref{sec:srgt} summarizes the basic notions from graph theory that
we use in the text.
 
-- Appendix \ref{ShortRemindMC} (resp. \ref{ShortRemindMP}) gather notions on
(time-homogeneous) finite state Markov chains (resp. Markov processes). We
stress that trajectories of Markov chains (resp. Markov processes) can be
analyzed as jumps governed by a Markov chain separated by independent geometric
(resp.  exponential) waiting times. This viewpoint is most useful for our
discussion and is often less familiar to physicists than the master equation
approach. We briefly mention the graph theoretic interpretation of transitions.

\vspace{.2cm}

There is an extensive literature dealing with the role of cycles appearing in
samples of Markov chains or processes with a countable number of states, see
e.g. \cite{jiang-qian-qian-2004,kalpazidou2006}. The fundamental role played by
cycles, affinity and entropy production to understand recurrence properties and the
structure of stationary measures has a number of applications to other fields
including deterministic dynamical systems. 

\section{Affinities and their invariances} \label{sec:ati}

This section is mostly for motivations. Our aim is mainly to recall the
fundamental role played by affinities in non-equilibrium systems.

Though the presentation itself is not totally standard, most results (in
particular those concerned with decompositions of entropy variations, with the
exception of \eqref{eq:poscur}) are.

\subsection{Cumulative processes} \label{ssec:cp}

Consider a Markov process $(\mathbf{C},\Trans)$ (where $\mathbf{C}$ is a finite
set of configurations and $\Trans$ is the matrix of transitions rates from one
configuration to another, see Appendix \ref{ShortRemindMP} for further notations
and definitions).

The trajectory $\C_t$, $t\in [0,+\infty[$, a random function from $[0,+\infty[$
to $\mathbf{C}$ where $\C_t$ denotes the configuration at time $t$, describes a
particle jumping from one configuration to another at certain instants and an
interesting class of observables just counts something at each jump. We can
describe this by introducing an arbitrary matrix $\Q$ on $\mathbf{C}$ with
vanishing diagonal elements. To $\Q$, we associate a process $\XQ=(\XQ_t)_{t\in
[0,+\infty[}$ defined as follows: $\XQ_0=0$, and each time there is a jump from a
configuration $\C$ to a configuration $\C'$, $\XQ$ jumps by $(\C'\vert \Q \vert
\C)$. Formally:
\[ \XQ_t\equiv \sum_{s\in ]0,t]} (\C_s\vert \Q \vert \C_{s^-}),\] so that indeed
$\XQ$ changes only when the particle jumps, i.e. for those $t$ such that $\C_t
\neq \C_{t^-}$. The sum is well-defined because on $]0,t]$ there are only
finitely many jumps with probability $1$ (this is a theorem, but the intuitive
reason is clear: on average, the time between two jumps is bounded below by the
inverse of the absolute value of the smallest diagonal element of the Markov
matrix). A generic name for this type of processes would be ``cumulative
processes''. They are a special case of the so-called ``additive functionals''
used in the mathematical literature.  A natural candidate for $\Q$ would be to
take all the non diagonal matrix elements equal to $1$. Then $\XQ$ would simply
count the number of jumps in $]0,t]$ -- what some authors call activity -- surely an important observable.

Note that the value of $(\C'
\vert \Q \vert \C)$ is immaterial if $(\C' \vert \Trans \vert \C)=0$. We could
for instance take the convention that $(\C' \vert \Q \vert \C)=0$ whenever $(\C'
\vert \Trans \vert \C)=0$. With this convention, $\Q$ would appear as a function
on edges of the graph $\mathbf{G}$ associated to $(\mathbf{C},\Trans)$. But
later, we shall give arguments to concentrate on anti-symmetric $\Q$'s, and this
might lead to conflicts with this convention.  Under the hypothesis of
micro-reversibility, which is our main interest, there is no such conflict.

Suppose the Markov process is known up to time $t$. What is $\XQ_{t+\Delta
  t}-\XQ_t$ in average? By definition, $\XQ_{t+\Delta t}-\XQ_t= \sum_{s\in
  ]t,t+\Delta t]} (\C_s\vert \Q \vert \C_{s^-})$. With probability $1-\Delta
t\sum_{\C' \neq \C_t} (\C' \vert \Trans \vert \C_t) + o(\Delta t)$ there has
been no jump in $]t,t+\Delta t]$, and with probability $\Delta t (\C' \vert
\Trans \vert \C_t) + o(\Delta t)$ there has been a jump to $\C' \neq C_t$. The
possibility of several jumps is negligible. So the average of $\XQ_{t+\Delta
  t}-\XQ_t$ if the process is known up to time $t$, which is in the
probabilistic language a conditional expectation\footnote{On conditional
  expectations, see the general references given at the beginning of Appendix
  \ref{ShortRemindMC}.} is
\[\Delta t\sum_{\C' \neq \C_t} (\C' \vert \Trans \vert \C_t) (\C' \vert \Q
\vert \C_t)+ o(\Delta t).\]
This leads to introduce
\be \label{eq:current} \jQ_t \equiv \sum_{\C' \neq \C_t} (\C' \vert \Trans \vert \C_t) (\C' \vert \Q
\vert \C_t)= (\C_t \vert \Q^{\dag} \Trans \vert \C_t),\ee where $\Q^{\dag}$
denotes the transpose of $\Q$.

Note that $\jQ_t=\jQ(\C_t)$ if $\jQ(\C)\equiv \sum_{\C' \neq \C} (\C' \vert
\Trans \vert \C) (\C' \vert \Q \vert \C)$. The function $\jQ(\C)$ (or the
process $\jQ_t$) is called the current associated to $\Q$. By construction,
$\int_0^t\jQ_s ds$ is a continuous process but, though the process $\XQ_t$ has
jumps, they have the same average: \be \label{eq:averxj} \Esp{\XQ_t} =
\Esp{\int_0^t\jQ_s ds}=\int_0^t \Esp{\jQ_s}ds,\ee where $\Esp{\cdot}$ denotes
the expectation or average with respect to the probability law $\Prob$ of the
Markov process with characteristics $(\mathbf{C},\Trans,\Pin)$ where $\Pin$ is
the initial probability distribution. The proof is simple. If
$\Esp{\XQ_{t+\Delta t}-\XQ_t\vert \text{ knowing }\C_s,s\in [0,t]}$ denotes the
average of $\XQ_{t+\Delta t}-\XQ_t$ when the process is known up to time $t$, we
have seen that
\[
\Esp{\XQ_{t+\Delta t}-\XQ_t\vert \text{ knowing }\C_s,s\in [0,t]}=\jQ_t \Delta t
+ o(\Delta t).\] On the left-hand side we have already averaged over the
fluctuations of the trajectory in the time interval $]t,t+\Delta t]$ so if we
average now over the possible trajectories $\C_s,s\in [0,t]$, we simply get
$\Esp{\XQ_{t+\Delta t}-\XQ_t}$. The right-hand side is insensitive to
fluctuations in the trajectory after $t$ so if we average now over the possible
trajectories $\C_s,s\in [0,t]$ we get $ \Esp{\jQ_t}\Delta t + o(\Delta t)$. So
$\Esp{\XQ_{t+\Delta t}-\XQ_t}= \Esp{\jQ_t}\Delta t + o(\Delta t)$. The equality
\eqref{eq:averxj} follows by taking Riemann sums and letting $\Delta t \to 0$.

Note that $\Esp{\jQ_s}=\sum_{\C,\C'} (\C' \vert \Trans \vert \C) (\C' \vert \Q
\vert \C)\Prob(\C;s)$, where $\Prob(\C;s)\equiv\Prob(\C_s=\C)$ is the
probability to be in configuration $\C$ at time $s$, and that $\int_0^t \Prob(\C;s) ds= \Esp{\int_0^t
  \mathbf{1}_{\C_s=\C} ds}$ is just the expectation
of the time spent at $\C$ in the interval $[0,t]$. Hence,  $\Esp{\XQ_t}$ can be
written in terms of occupation times. 

Though this does not play any role in the sequel, let us stress that the
relation between a cumulative process $\XQ$ and the associated current $\jQ$ has
a deep probabilistic meaning\footnote{In proper mathematical language, $\XQ$ is
  a special semi-martingale, and its unique decomposition
  $\XQ_t=\Big(\XQ_t-\int_0^t\jQ_s ds\Big)+ \int_0^t\jQ_s ds$ as the sum of a
  martingale and a predictable finite variation process contains more than the
  mere coincidence $\Esp{\XQ_t}=\Esp{\int_0^t\jQ_s ds}$. Though powers of $\XQ$
  are still special semi-martingales, their decomposition as a martingale plus a
  predictable finite variation process involves other currents. The systematics
  of this decomposition, which will be given elsewhere, is physically relevant.
  Indeed, in a number of experiments, the quantity of interest is some $\XQ$,
  but only $\jQ$ and possibly a few other currents are measurable. The
  information carried by these currents on fluctuations of $\XQ$ (as embodied in
  moments $\Esp{\big(\XQ_t\big)^k}$ for instance) is limited and some
  combinatorial effort is required to extract it.}. We hope to return to this in
a forthcoming work \cite{bauer-cornu2014}.

\subsection{Exchange processes} \label{ssec:ep}

In the sequel, we shall concentrate on \textit{anti-symmetric} $\Q$'s (which is
not the case when one counts jumps). The physical reason is the following: we
view the jumps in the Markov process as triggered by interactions with some
external reservoirs, leading to the exchange of some conserved quantities
(energy, particles,...). The reservoir responsible for a jump from $\C$ to $\C'$
would give to the system a quantity $(\C' \vert \Q \vert \C)$ which should be
given back during the reverse jump from $\C'$ to $\C$. Note that this is
really a constraint only if $(\C' \vert \Trans \vert \C)$ and $(\C \vert \Trans
\vert \C')$ are both non-vanishing. When $\Q$ is anti-symmetric, we call $
\XQ_t$ the exchange process, and $\jQ_t$ the exchange current, associated to
$\Q$. Note that we can rewrite $\jQ_t$ as a matrix product $\jQ_t=-(\C_t \vert
\Q \Trans \vert \C_t)$.

Another reason for the physical relevance of anti-symmetric $\Q's$ is that
detailed balance is easily formulated: suppose that $\Pit$ is a probability on
$\mathbf{C}$. The average of $\jQ$ under $\Pit$, $\Esp{\jQ}_\Pit \equiv
\sum_{\C}\jQ(\C) \Pit(\C)$ vanishes for every anti-symmetric $\Q$ if and only if
$(\C' \vert \Trans \vert \C)\Pit(\C)= (\C \vert \Trans \vert \C')\Pit(\C')$ for
every $\C, \C' \in \mathbf{C}$. As usual, this implies that $\Pit$ is a
stationary measure for $(\mathbf{C},\Trans)$. Hence, if $\Prob(\cdot;t)$ has a
limit when $t \to +\infty$, this limit satisfies detailed balance if and only if
the average of every exchange current $\Esp{\jQ_t}$ goes to $0$ when $t \to
+\infty$, and then so does $\frac{1}{t}\Esp{\XQ_t}=\frac{1}{t}\int_0^t
\Esp{\jQ_s} ds$. Note that we use the name ``detailed balance'' in a slightly
generalized sense, because we have not specified an energy
function on $\mathbf{C}$, defined a corresponding Boltzmann weight and so on.

There is also a mathematical reason, related to (co)homology, to focus on
anti-symmetric $\Q's$. Assuming micro-reversibility, we view anti-symmetric
$\Q$'s as functions on edges of the graph $\mathbf{G}$, changing sign when the
edge orientation is reversed. By definition, these are just the \textit{$1$-cocycles} on
$\mathbf{G}$. Special $1$-cocycles are \textit{$1$-coboundaries}, when $(\C' \vert \Q
\vert \C)=\Oobs(\C')-\Oobs(\C)$ for some observable $\Oobs$, i.e. for some
function on $\mathbf{C}$. Notice that in that case
$\XQ_t=\Oobs(\C_t)-\Oobs(\C_0)$ depends only on the configurations at time $t$
and time $0$, but that nevertheless the compensator $\int_0^t \Esp{\jQ_s} ds$
depends on the whole past. A sequence $(\C_0,\C_1,\cdots,\C_n)$ of vertices of
$\mathbf{G}$ is called a walk (of length $n$) on $\mathbf{G}$ if
$(C_m,\C_{m+1})$ is an edge of $\mathbf{C}$ for $0 \leq m < n$. By construction,
a trajectory of the Markov process $(\mathbf{C},\Trans)$ up to (and including)
the $n^{\text{th}}$ jump is a walk of length $n$ on $\mathbf{G}$. A walk with
$n\geq 3$ and $\C_n=\C_0$ is called a \textit{$1$-cycle} of $\mathbf{G}$. One can
``integrate'' a $1$-cocycle along a walk : $\intp_{(\C_0,\C_1,\cdots,\C_n)} \Q
\equiv \sum_{m=0}^{n-1} (\C_{m+1} \vert \Q \vert \C_m)$ (the notation $\intp$ is
not standard, it is introduced here just to stress the formal analogy with more
familiar line integrals). The process $\XQ_t$ is just the ``integral'' of $\Q$
along the walk described by the trajectory up to time $t$.

The ``integral'' of a $1$-coboundary along a $1$-cycle is always $0$, and one can
prove that conversely, if the integral of a $1$-cocycle along any $1$-cycle of
$\mathbf{G}$ vanishes, then the $1$-cocycle is a $1$-coboundary. The
``integral'' of a $1$-coboundary along a path depends only on the origin and end
of the path. If the $1$-cocycle $\Q$ is in fact a hidden $1$-coboundary, one can
recover the corresponding observable by choosing one ``base'' configuration in
each connected component of $\mathbf{G}$, and setting $\Oobs(\C) \equiv
\intp_{\text{Any path from a base configuration to } \C} \Q$. Of course $\Oobs$
is only defined up to arbitrary additive constants (one for each connected
component). 

One can view $1$-cycles as analogs of thermodynamic cycles, and $1$-coboundaries
as variations of state functions, while $1$-cocycles are in some sense analogs of
heat exchanges.

\subsection{Affinities} \label{ssec:aff}

In this section, where we assume micro-reversibility (see Appendix
\ref{sec:srgt} and \ref{ShortRemindMP} for reference) all along, we come to
particular exchange processes: those which can be defined solely in terms of the
basic data of the Markov process $(\mathbf{C},\Trans)$. Namely, we have at our
disposal $\Trans$, and possibly a probability distribution $\Pit$ on
$\mathbf{C}$, or the time evolved probability distribution $\Prob(\C;t)$.

The first exchange that comes to mind is probably $\Trans-\Trans^{\dag}$, but is
does not play an important role in the sequel. 

The next one in terms of complexity is perhaps $\mathbb{S}$ defined by
\[ (\C'\vert \mathbb{S} \vert \C) \equiv \ln \frac{(\C'\vert \Trans\vert \C)}
{(\C\vert \Trans\vert \C')}\text{ if } \{\C,\C' \} \text{ is an edge of }
\mathbf{G}\text{ and } 0 \text{ if it isn't}. \]The quantity $\mathbb{S}$ is
appealing for a
number of reasons.\\
-- It is dimensionless.\\
-- The corresponding process $\XS_t$ is exactly the Lebowitz-Spohn action
functional \cite{LebowitzSpohn1999} for the trajectory up to time $t$.  From a
purely Markov process viewpoint, $e^{\XS_t}$ is, up to boundary terms, nothing
but the ratio of the weight of the trajectory $\C_s$, $s\in [0,t]$ and its time
reversal
$\C_{t-s}$ $s\in [0,t]$ (a Radon-Nykodim derivative to be precise). \\
-- The statement ``The $1$-cocycle $\mathbb{S}$ for $(\mathbf{C},\Trans)$ is a
$1$-coboundary'' (for which it is enough to check that the ``integral''
$\intp_{\text{cycle}} \mathbb{S}$ vanishes for every cycle of $\mathbf{G}$) is
an elegant mathematical way to say ``Detailed balance is satisfied for
$(\mathbf{C},\Trans)$''. Indeed, if $\mathbb{S}$ is a $1$-coboundary we can
write $(\C'\vert \mathbb{S} \vert \C)=\ln w(\C')-\ln w(\C)$ for a strictly
positive function $w$ on $\mathbf{C}$. This means that $\frac{(\C'\vert
  \Trans\vert \C)} {(\C\vert \Trans\vert \C')}=\frac{w(\C')}{w(\C)}$ for each
edge of $ \mathbf{G}$.  Multiplying $w$ by a function which is constant on each
connected component of $\mathbf{G}$ leaves this relation invariant. In
particular, one may assume that $w$
is a probability (i.e. $\sum_{\C} w(\C)=1$), leading to detailed balance. \\
-- Finally, we  emphasize that anti-symmetry is related to an
interpretation of exchanges with reservoirs, and simple physical models   \cite{Schnakenberg1976,Derrida2007,CornuBauerAPourBETA} support
the interpretation that $(\C'\vert \mathbb{S} \vert \C)$ is just the variation
of entropy of the reservoirs when the system transits from $\C$ to $\C'$.

\vspace{.2cm}

Our next aim is to recall a standard entropy argument, which is essentially
already in \cite{Schnakenberg1976}, supporting this view.

\subsubsection{Entropy variations and chemical affinity}

The first observation is that $\jS (\C)= \sum_{\C' \neq \C} (\C' \vert \Trans
\vert \C) (\C' \vert \mathbb{S} \vert \C)$, so that $\Esp{\jS_t}=\sum_{\C, \C'}
(\C' \vert \Trans \vert \C) (\C' \vert \mathbb{S} \vert \C)\Prob(\C;t)$ which by
anti-symmetry can be rewritten as
\[\Esp{\jS_t}=\frac{1}{2}\sum_{\C, \C'} \dot{\Extent}_{\C;\C'}\ln \frac{(\C'\vert
  \Trans\vert \C)} {(\C\vert \Trans\vert \C')},\] where we have set
$\dot{\Extent}_{\C;\C'}\equiv -\dot{\Prob}_{\C;\C'}(t)=-(\C \vert \Trans \vert
\C')\Prob(\C';t)+(\C' \vert \Trans \vert
\C)\Prob(\C;t)$. The rationale for this notation is that:\\
-- The equation for the time evolution of $\Prob(\C;t)$ can be rewritten as
\[\frac{d\Prob(\C;t)}{dt}=\sum_{\C'} \dot{\Prob}_{\C;\C'}(t),\]
i.e. $\dot{\Prob}_{\C;\C'}(t)$ is the contribution of edge $\{\C, \C'\}$ to the
variation of $\Prob(\C;t)$ with time. \\
-- Consequently, viewing $\C,\C'$ as chemical species and the edge $\{\C, \C'\}$
as a chemical reaction, \[\int_0^t \dot{\Extent}_{\C;\C'}(s)ds=-\int_0^t
\dot{\Prob}_{\C;\C'}(s)ds \] is by definition the ``extent of reaction'' (modulo
a substitution of probabilities for concentrations), quantifying how much the
reaction $\C \leftrightarrow \C'$ has progressed between time $0$ and $t$. The
minus sign is because the extent of the reaction grows when the reactant
concentration drops down.

The next step is to introduce the (dimensionless) Shannon-Gibbs entropy at time
$t$. Recall that if $\mu$ is any probability distribution on $\mathbf{C}$, then
$\SG[\mu]\equiv - \sum_{\C}\mu(\C)\ln\mu(\C)=-\Esp{\ln \mu}_{\mu}$ where
$\Esp{\cdot}_\mu$ is the expectation for a function on $\mathbf{C}$ under the
probability distribution $\mu$, not to be confused with $\Esp{\cdot}$ which denotes
in this paper the expectation with respect to the Markov process probability
measure on paths\footnote{There is a relationship though. If $\Oobs$ is an
  observable, i.e.  a function on $\mathbf{C}$, one can associate to it a random
  variable depending on the path, $\Oobs(\C_t)$. The path dependence is only via
  the configuration $\C_t$ at time $t$, and in that case the expectation with
  respect to the probability measure on paths can be computed by taking an
  expectation with respect to its one-time marginal $\Prob(\C,t)$ which is a
  probability on $\mathbf{C}$: $\Esp{\Oobs(\C_t)}=\Esp{\Oobs}_{\Prob(t)}$.}.
The entropy of the probability distribution $\Prob(\C;t)$ is thus \be
\label{defSG}
\SG[\Prob(t)]=- \sum_{\C}\Prob(\C;t)\ln\Prob(\C;t).
\ee
Note that Boltzmann's constant is set equal to $1$, hence the name
``dimensionless''. 

Simple manipulations show that 
\[
\frac{d\SG}{dt}=\frac{1}{2}\sum_{\C,\C'} \dot{\Extent}_{\C;\C'}(t)\ln \frac{\Prob(\C;t)}{\Prob(\C';t)}
\]
which can be split in two:
\[
\frac{d\SG}{dt}= \frac{1}{2}\sum_{\C,\C'} \dot{\Extent}_{\C;\C'}(t)\ln \frac{(\C' \vert \Trans \vert \C)\Prob(\C;t)}{(\C \vert \Trans \vert
\C')\Prob(\C';t)}-\Esp{\jS_t}
\]
Each summand in the first term is $\geq 0$ (it is of the form $-(x'-x)\log
(x/x')$), so it contributes to a systematic increase of entropy, and can be
interpreted as a sign of irreversibility of the chemical reaction $\C
\leftrightarrow \C'$. Then each summand in the second term can be interpreted as
the entropy change due to exchanges with reservoirs, thus recovering the usual
splitting of entropy variations as the sum of two contributions: one which is
always positive and signs irreversibility, and one leading to entropy variations
in the reservoirs: $d\SG= (d\SG)_\text{\mdseries irr} + (d \SG)_\text{\mdseries
  exch}$. Note that the interpretation of $e^{\XS_t}$ given at the beginning of
Section \ref{ssec:aff} as related to time reversal supports strongly that $(d
\SG)_\text{\mdseries exch}$ is a reversible contribution. The standard name in
the literature for the irreversible contribution $(d\SG)_\text{\mdseries irr}$
is simply ``entropy production'' and its crucial role to understand
probabilistic properties (for instance recurrence) is widely recognized (see
e.g. \cite{Schnakenberg1976,LebowitzSpohn1999,jiang-qian-qian-2004,kalpazidou2006,MaesNetocnyWynants2008}).

The above discussion is only one of the lines of reasoning leading to this
decomposition (see e.g.
\cite{MaesNetocnyWynants2008,Muratore-Ginanneschi-et-al-2013}). In certain
circumstances, when exchanges with reservoirs are explicitly built in the
transition rates (see e.g.  \cite{Derrida2007,CornuBauerAPourBETA}), there is a
direct computation at the level of the reservoirs leading to the equality (not
just an abstract interpretation) $(d \SG)_\text{\mdseries
  exch}=-dS^{\scriptscriptstyle Res}$, where $dS^{\scriptscriptstyle Res}$ is
the entropy variation of the reservoirs due to exchanges with the system. Let us
stress however that even if these arguments, analogies and identifications are
convincing and well-motivated, their universal validity for out-of-equilibrium
systems is not proven.

Returning to the main stream of the argument, we thus have 
\[ (d\SG)_\text{\mdseries irr} = \frac{1}{2}\sum_{\C,\C'} \dot{\Extent}_{\C;\C'}(t)\ln \frac{(\C' \vert \Trans \vert \C)\Prob(\C;t)}{(\C \vert \Trans \vert
\C')\Prob(\C';t)} \, dt  \qquad  (d \SG)_\text{\mdseries exch} = -\Esp{\jS_t}\, dt.\]

Finally, we use the chemical definition of affinity: if $\xi$ is the ``extent of
a reaction'' and $S$ is the entropy, the affinity $A$ is defined\footnote{ The
  term ``affinity'' was coined by De Donder (see e.g. \cite{Prigogine1968} for an
  introduction in English).} via $ (dS)_\text{\mdseries irr}= A d\xi$. For the
reaction $\C \leftrightarrow \C'$, we have seen that
$d\xi=\dot{\Extent}_{\C;\C'}dt$. Putting all these considerations together, we
obtain that the affinity of the reaction $\C \leftrightarrow \C'$ is $\ln
\frac{(\C' \vert \Trans \vert \C)\Prob(\C;t)}{(\C \vert \Trans \vert
  \C')\Prob(\C';t)}$ (there is no $1/2$ because $\sum_{\C,\C'}$ counts every
unoriented edge, i.e. every reaction, twice).

If $\Pit$ is a probability distribution on $\mathbf{C}$, we define $\Aev^{\Pit}$
by 
\[(\C' \vert \Aev^{\Pit} \vert \C)=\ln \frac{(\C' \vert \Trans \vert
  \C)\Pit(\C)}{(\C \vert \Trans \vert \C')\Pit(\C')}, \] so that $(\C' \vert
\Aev^{\Pit} \vert \C)$ is the affinity of the reaction $\C \leftrightarrow \C'$
when the probability distribution of the system is $\Pit$. For fixed $\Pit$,
$\Aev^{\Pit}$ is an exchange, or $1$-cocycle, and if the system is in state
$\Pit$ we can write
\[ (d\SG)_\text{\mdseries irr} =\Esp{j^{\Aev^{\Pit}}}_\Pit \,dt.\]

The fact, recalled above, that $\Esp{j^{\Aev^{\Pit}}}_\Pit$ is always $\geq 0$ is
well-documented in the literature. However, we have not found there that for
$\Pit=\Probst$ (the stationary measure) the current $j^{\Aev^{\Probst}}$ itself
is positive: 
\be \label{eq:poscur}
j^{\Aev^{\Probst}}(\C)=\sum_{\C'} (\C' \vert \Trans \vert \C) \ln \frac{(\C'
  \vert \Trans \vert \C)\Probst(\C)}{(\C \vert \Trans \vert \C')\Probst(\C')}\,
\geq 0 \; \forall\, \C .
\ee 
This fact can be proven by hand using $- \ln x \geq
1-x$ for $x \geq 0$ and the definition of $\Probst$: $\sum_{\C'}(\C\vert \Trans\vert \C')
\Probst(\C')=\sum_{\C'}(\C'\vert \Trans\vert \C) \Probst(\C)$. It is also a
consequence of the interpretation of $j^{\Aev^{\Probst}}(\C)$ for each fixed
$\C$, rewritten as \[-\sum_{\C'} (\C' \vert \Trans \vert \C) \ln
\frac{\Probst(\C)^{-1}(\C \vert \Trans \vert \C')\Probst(\C')} {(\C' \vert
  \Trans \vert \C)}\] i.e. as a relative entropy.

One can carry a similar analysis for Markov chains: if $(\mathbf{C},\Pdt)$ is a
Markov chain, and under the usual assumption of micro-reversibility, one can
show that the (discrete time) variation of the entropy leads to consider 
\be \label{eq:discr-aff} \ln \frac{(\C'\vert \Pdt \vert \C)}
  {(\C\vert \Pdt \vert \C')} \text{ and } \ln \frac{(\C' \vert \Pdt \vert
    \C)\Pit(\C)}{(\C \vert \Pdt \vert \C')\Pit(\C')}
\ee
 as the
discrete time analogs of $\mathbb{S}$ and $\Aev^{\Pit}$.

Note that $\Aev^{\Pit}$ is a $1$-cocycle and that $\Aev^{\Pit}-\mathbb{S}$ is a
$1$-coboundary, as already noticed by Schnackenberg in a different language
\cite{Schnakenberg1976}. So the ``integrals'' of $\Aev^{\Pit}$ and $\mathbb{S}$
along any cycle of $\mathbf{G}$ agree. As our interest in the sequel is mainly
in such ``integrals'', there is no reason to consider $\Aev^{\Pit}$ and
$\mathbb{S}$ as different objects. Another way to phrase this idea,
maybe more familiar for physicists, is to view $1$-coboundaries as
gauge transformations in an abelian gauge theory, as emphasized in
\cite{polettini2012} as we learned after this work was completed.

In the (co)homology language, one regroups all $1$-cocycles differing by a
$1$-coboundary under the name ``cohomology class''. All we shall keep from the
concept is the name. So we define the affinity class as the class of all
exchanges that differ from $\mathbb{S}$ by a $1$-coboundary. As seen above, this
class comprise $\Aev^{\Pit}$ whatever $\Pit$.  

The above symmetry is somewhat formal at the moment, and our aim in the next
section is to show that the affinity class is preserved by many natural
probabilistic constructions.

\section{Probabilistic constructions}

\label{sec:acpc}

\subsection{A ``graphical'' construction: Drag and drop} \label{ssec:dd} 

At the basis of our interpretation of affinities lies a simple\footnote{So
  simple that it is likely to have already been introduced elsewhere, but we are
  not aware of a reference.} general construction on (oriented or non-oriented)
graphs.  We refer the reader to Appendix \ref{sec:srgt} for basic definitions
and notations for graphs. This construction has several variants.  Suppose that
$\mathbf{G}$ is a graph and $\mathbf{H}$ a subgraph of $\mathbf{G}$.\\
\underline{First variant:} Take a walk $X=(X_0,\cdots,X_N)$ on $\mathbf{G}$ and
let $Y_0$ be a vertex of $\mathbf{H}$. Construct recursively a sequence
$Y=(Y_0,\cdots,Y_N)$ of vertices of $\mathbf{H}$ as follows. The first term
$Y_0$ is already defined. Suppose that, for some $0 \leq n < N$,
$Y_0,\cdots,Y_n$ are already defined. If $X_n=Y_n$ and $(X_n,X_{n+1})$ is an
edge of $\mathbf{H}$ then set $Y_{n+1}=X_{n+1}$, otherwise, set $Y_{n+1}=Y_n$. By
construction, $Y_0,\cdots,Y_N$ is a sequence of vertices of $\mathbf{H}$, but in
general is does not have to be a walk on $\mathbf{H}$ because $Y$ can stay at
the same place even if there can be no edge having this vertex both as
origin and extremity.\\
\underline{Second variant:} Take a walk $X=(X_0,\cdots,X_N)$ on $\mathbf{G}$ and
let $Y_0$ be a vertex of $\mathbf{H}$. Construct recursively a walk
$Y=(Y_0,\cdots,Y_M)$ on $\mathbf{H}$ and a sequence $n_0=0<n_1<\cdots <n_M\leq
N$ for some $0 \leq M \leq N$ as follows.  The first terms $Y_0$ and $n_0$ are
already defined.  Suppose that for some $0 \leq m$, $Y_0,\cdots,Y_m$ and
$n_0=0<n_1<\cdots <n_m$ are already defined. If it exists, let $n$ be the
smallest integer $n_m \leq n < N$ such that $X_n=Y_m$ and $(X_n,X_{n+1})$ is an
edge of $\mathbf{H}$ and set $Y_{m+1}=X_{n+1}$, $n_{m+1}= n+1$. If no such $n$
exists, set $M=m$ and stop.

Of course, we could replace the finite walk $X_0,\cdots,X_N$ by an infinite walk
$X_0,X_1,\cdots$ and do the same construction. The second variant is most useful
for our analysis. By lack of a better name, we choose to call it ``drag and
drop'' along $\mathbf{H}$.

The intuitive picture is that $Y$ represents the position of an object (a box)
attached to $\mathbf{H}$ and the walker $X$ carries $Y$ along
$\mathbf{H}$ when they meet and $X$ moves along $\mathbf{H}$. The second
variant is closely related to the first one. In the first variant, $Y$
and $X$ are indexed by the same ``clock'', in the second variant, the
clock of $Y$ ticks only when $X$ carries $Y$ along
$\mathbf{H}$.

There is a special case of ``drag and drop'' which is already of some interest.
It happens when all walks on $\mathbf{G}$ joining two distinct vertices of
$\mathbf{H}$ involve at least one edge of $\mathbf{H}$\footnote{For
  microreversible graphs, this can be stated as follows: if the edges of
  $\mathbf{H}$ are removed from $\mathbf{G}$, distinct vertices of $\mathbf{H}$
  belong to disjoint connected component of the resulting graph.}. In that case,
the box is not really needed: to go from the walk $X=(X_0,\cdots,X_N)$ to the
walk $Y=(Y_0,\cdots,Y_M)$ first remove all vertices up to the first appearance
of $Y_0$, then all vertices after $Y_0$ that do not belong to $\mathbf{H}$, and
finally remove multiple contiguous occurrences of one and the same vertex.  This
simple special case is relevant when joining two distinct vertices of
$\mathbf{H}$ without going through an edge of $\mathbf{H}$ requires rather long
walks, which can be neglected to a good approximation in practice. This could be
used for instance to get a crude model of the experiment in \cite{DiLeonardo-et-al}.

It is not unlikely that this kind of coupled motion is routinely performed by
the biological molecular machinery, though the microscopic understanding is
still too preliminary to know for sure \cite{mcmk-private}.\\
In such a scenario, $\mathbf{H}$ models for instance some hetero-polymer
within a cell and $\mathbf{G}$ models $\mathbf{H}$ plus the relevant environment
of $\mathbf{H}$ inside the cell.  There is a ligand bound at some place on
$\mathbf{H}$ but that may move along it. There is a complex that moves on
$\mathbf{G}$. When it meets the ligand somewhere on $\mathbf{H}$, it binds to
the ligand. If the complex moves along $\mathbf{H}$ it carries the ligand, but
if it goes somewhere else the ligand and complex unbind and the ligand waits for
another encounter with the complex.\\
By the way, in this interpretation, there may be several complexes that cannot
occupy the same position at the same time and it is trivial to generalize the
above construction to this more general case. Then, the process becomes close to
hitch-hiking. The hitch-hiker wants to go somewhere or make some journey, but
several roads may be acceptable for him. When a car stops and the driver offers a
lift to a place that is compatible with the hitch-hiker's goals, they do a piece
of journey together after which the driver drops the hitch-hiker and drives his
way, while the hitch-hiker waits for another driver to continue his journey.

This generalization is also appropriate to make a connection with some recent
experiments in active matter with applications to nanotechnologies, see e.g.
Ref.\cite{DiLeonardo-et-al} entitled ``Bacterial Ratchet Motors'', for which the
special case of ``drag and drop'' is already of some relevance.

\subsection{Drag and drop for Markov processes} \label{ssec:ddmp}

We would like to understand ``drag and drop'' when the walker moves according to
a finite state Markov process. We refer the reader to Appendix
\ref{ShortRemindMP} for the relevant notations and basic constructions. To avoid
some trivial situations, we assume that the number of configurations is $\geq
2$. We denote by $\Trans$ the transition matrix and by $\M=\Trans-\D$ the
infinitesimal generator of a finite state Markov process, by $\Pdt$ the
associated stochastic matrix satisfying $\M=(\Pdt-\Id)\D$ and by $\mathbf{G}$
the associated graph. We let $\mathbf{H}$ be a subgraph of $\mathbf{G}$ and
denote by $\Trans_{\mathbf{H}}^{\text{rest}}$ the restriction of the transition
matrix to $\mathbf{H}$, i.e. $(\C'\vert \Trans_{\mathbf{H}}^{\text{rest}} \vert
\C)$ is defined only when $\C,\C'$ are vertices of $\mathbf{H}$ and then:
$(\C'\vert \Trans_{\mathbf{H}}^{\text{rest}} \vert \C)\equiv (\C'\vert \Trans
\vert \C)$ if $(\C,\C')$ is an edge of $\mathbf{H}$ and $0$ otherwise.  From
$\Trans_{\mathbf{H}}^{\text{rest}}$ viewed as a transition matrix (indexed by
vertices of $\mathbf{H}$) we construct as recalled in Appendix
\ref{ShortRemindMP} an infinitesimal generator $\M_{\mathbf{H}}^{\text{rest}}=
\Trans_{\mathbf{H}}^{\text{rest}}- \D_{\mathbf{H}}^{\text{rest}}$ and a
stochastic matrix $\Pdt_{\mathbf{H}}^{\text{dd}}$ (where the superscript
$\text{dd}$ stands for ``drag and drop'') satisfying
$\M_{\mathbf{H}}^{\text{rest}}=(\Pdt_{\mathbf{H}}^{\text{dd}}-
\Id_{\mathbf{H}})\D_{\mathbf{H}}^{\text{rest}}$.

We can apply the general construction of ``drag and drop''.  Take a trajectory
of the Markov process on $\mathbf{G}$ with infinitesimal generator $\M$ and
initial probability distribution $\Prob(\C,0)$.  If $0,T,T+T',\cdots$ denotes
the (finite or infinite) sequence of jump times we let $(\C,\C',\C'',\cdots)$
denote the (finite or infinite) sequence of corresponding positions on the
graph: the process is at $\C$ between $0$ and $T$, at $\C'$ between $T$ and
$T+T'$, and so on. Keeping in mind our remarks (see Appendix
\ref{ShortRemindMP}) on the connection between a
finite state Markov process and the associated finite state Markov chain, if the
sequence $(\C,\C',\C'',\cdots)$ is finite, we turn it into an infinite one by
repeating the last member again and again. Thus we have the first ingredient for
``drag and drop''.  The second ingredient is the initial configuration
$\bar{\C}$ on $\mathbf{H}$ that will be ``dragged and dropped'' by the walk
$(\C,\C',\C'',\cdots)$. Applying the \textit{second} variant of ``drag and
drop'' we get a (finite or infinite) sequence
$(\bar{\C},\bar{\C}',\bar{\C}'',\cdots)$ on $\mathbf{H}$ and a (finite or
infinite) sequence $0,\bar{T}, \bar{T}+ \bar{T}',\cdots$ of times at which the
jumps along $\mathbf{H}$ occur. Note that these sequences may be finite even
though the sequence $(\C,\C',\C'',\cdots)$ is infinite. In this way, we have
defined a (continuous time) random process with the vertices of $\mathbf{H}$ as
configurations. 

The first question that comes to mind is whether this process is Markov, and a
moment thinking shows that the answer is no in general. Indeed, the waiting time
between two jumps along $\mathbf{H}$ is in general not exponential: it is a
complicated mixture (we use the term in an informal way) of the exponential
waiting times of the original Markov process.

However, some remnant of the Markov property is preserved. To state it properly,
we apply our usual trick to the sequence of successive positions on
$\mathbf{H}$: if the sequence $(\bar{\C},\bar{\C}',\bar{\C}'',\cdots)$ is
finite, we turn it into an infinite one by repeating the last member again and
again. We have the following (we do not try to give minimal hypotheses).

\vspace{.2cm}

\textit{Claim}\\
Suppose that the initial point $\bar{\C}$ on $\mathbf{H}$ is recurrent for the
Markov process on $\mathbf{G}$, i.e. with probability $1$ configuration
$\bar{\C}$ appears infinitely many times in the sequence $(\C,\C',\C'',\cdots)$.
Then $(\bar{\C},\bar{\C}',\bar{\C}'',\cdots)$ is a Markov chain
started at $\bar{\C}$ with stochastic matrix $\Pdt_{\mathbf{H}}^{\text{dd}}$.

\vspace{.2cm}

\textit{Remark}\\
The hypothesis that $\bar{\C}$ on $\mathbf{H}$ is recurrent for the Markov
process on $\mathbf{G}$ is always fulfilled if micro-reversibility holds (in
which case non-oriented graphs are the right thing to look at) and $\mathbf{G}$
is connected.

\vspace{.2cm}

\textit{Sketch of proof}\\
By hypothesis $\bar{\C}$ is visited infinitely many times by the Markov process
on $\mathbf{G}$. \\
-- Either $\bar{\C}$ is not the origin of an outgoing edge of $\mathbf{H}$, and
then the sequence of positions on $\mathbf{H}$ is just the repetition of
$\bar{\C}$ forever, which is indeed a Markov chain started at $\bar{\C}$ because
$(\bar{\C}\vert \Pdt_{\mathbf{H}}^{\text{dd}}\vert\bar{\C})=1$,\\
-- Or let $(\hat{\C},\hat{\C}',\hat{\C}'',\cdots)$ be the sequence of vertices
of $\mathbf{G}$ visited by jumping from $\bar{\C}$. These vertices are chosen
independently and according to the probability measure $(\cdot \vert \Pdt\vert
\bar{\C})$ on vertices of $\mathbf{G}$. The first time one of the jumps is along
an edge of $\mathbf{H}$, this edge is sampled from the probability law $(\cdot
\vert \Pdt\vert \bar{\C})$ conditioned to $\mathbf{H}$ (see footnote
\footnote{If $K,K',K''\cdots$ is a sequence of independent identically
  distributed random elements with law $\mu$, then the first element of the
  sequence that belongs to some measurable set $B$ with $\mu(B) >0$ has law
  $\mu(\cdot |B)$, i.e. $\mu$ conditioned on $B$.  Indeed, if $A \subset B$ is
  measurable the event that the first term belonging to $B$ in fact belongs to
  $A$ is the disjoint union $\{K\in A\}\cup\{K\notin B, K'\in A\}\cup\{K\notin
  B, K'\notin B, K''\in A\}\cup \cdots$ which by independence has probability
  $\mu(A)+(1-\mu(B))\mu(A)+ (1-\mu(B))^2\mu(A)+\cdots= \mu(A)/\mu(B)$.}), which
is easily seen to be nothing but the probability law $(\cdot \vert
\Pdt_{\mathbf{H}}^{\text{dd}}\vert\bar{\C})$. Let $\bar{\C}'$ be the end of the
chosen edge.  Now if $\bar{\C}$ is visited infinitely many times by the Markov
process on $\mathbf{G}$, then so must be $\bar{\C}'$ and we can iterate the
argument, which works whether or not $\bar{\C}'$ is the origin of an outgoing
edge of $\mathbf{H}$.
\\
To conclude, $(\bar{\C},\bar{\C}',\bar{\C}'',\cdots)$ is a Markov chain started
at $\bar{\C}$ with transition matrix $\Pdt_{\mathbf{H}}^{\text{dd}}$.  \hfill
$\square$

As for the times $\bar{T},\bar{T}',\cdots$ between jumps along edges of
$\mathbf{H}$, we content with an informal discussion. As already stated, they
are not exponential in general (and their laws are quite complicated even for
$\mathbf{G}$ and $\mathbf{H}$ of modest size). But deep independence properties
survive.  

\vspace{.2cm}

\textit{Claim}\\
The sequence
$\Big((\bar{\C},\bar{T}),(\bar{\C}',\bar{T}'),(\bar{\C}'',\bar{T}'')
\cdots\Big)$, assumed to be infinite, is a Markov chain on the infinite
configuration space $\mathbf{C}\times ]0,+\infty[$ with a stochastic
matrix of a very special form (observe that the transition matrix factorizes and
contains no $\bar{T}$-dependence): 
\[ \Big( \bar{\C}',\bar{T}' \Big| \mbox{\Large $\Pdt$} \Big|
\bar{\C},\bar{T}\Big) = ( \bar{\C}' \vert \Pdt_{\mathbf{H}}^{\text{dd}} \vert
\bar{\C} ) \rho_{\bar{\C}'}(\bar{T}')\] where $\rho_{\cdot}(t)$ is a family of
probability densities on $]0,+\infty[$ indexed by the vertices of $\mathbf{H}$.

\vspace{.2cm}

This formulation is not fully correct, and without aiming at full rigour, we can
be a bit more formal: 

\vspace{.2cm}

\textit{Claim}\\
The sequence of pairs
$\Big((\bar{\C},\bar{T}),(\bar{\C}',\bar{T}'),(\bar{\C}'',\bar{T}'')
\cdots\Big)$ is a \textit{generalized renewal
  sequence}. 

\vspace{.2cm}

\textit{Remark}\\
By \textit{generalized renewal sequence} we mean the following.  We are given a
finite set of configurations (here, the vertices of $\mathbf{H}$), the stochastic matrix of some Markov
chain    on these configurations, (here,
$\Pdt_{\mathbf{H}}^{\text{dd}}$) and laws $\nu_{.},\mu_{.}$ (two for each
configuration i.e. here, two for each vertex of $\mathbf{H}$) for random
variables with values in $]0,+\infty[$. Then starting from a configuration
$\bar{\C}$ (which can be chosen randomly), one waits a time $\bar{T}$
distributed according to $\nu_{\bar{\C}}$ and jumps to another configuration
$\bar{\C}'$ with probability $(\bar{\C}' \vert
\Pdt_{\mathbf{H}}^{\text{dd}}\vert\bar{\C})$. Then independently of what
happened before (this is renewal) one waits at $\bar{\C}'$ a time $\bar{T}'$
distributed according to $\mu_{\bar{\C}'}$ and jumps to another configuration
$\bar{\C}''$ with probability $( \bar{\C}''\vert
\Pdt_{\mathbf{H}}^{\text{dd}}\vert\bar{\C}')$. Then independently of what
happened before one waits at $\bar{\C}''$ a time $\bar{T}''$ distributed
according to $\mu_{\bar{\C}''}$ and so on. Note that in general, the waiting
time at the initial configuration (law $\nu$) does not have to be distributed as
the waiting time at further passages at the same configuration (law $\mu$). In
our case, this is relevant because we do not impose that the initial
configuration $\C$ of the original Markov process on $\mathbf{G}$ be $\bar{\C}$
so $\bar{T}$ is the sum of two (independent) contributions: some time is spent
to reach $\bar{\C}$ from $\C$, and then a time distributed according to
$\mu_{\bar{\C}}$ is spent before the jump to $\bar{\C}'$, so $\nu_{\bar{\C}}$ is
a composite object. But for all further passages at $\bar{\C}$, the waiting
times to jump from $\bar{\C}$ will have law $\mu_{\bar{\C}}$. Note that the
density $\rho_{\bar{\C}}$ above is the density of the law $\mu_{\bar{\C}}$. Note
also that Markov processes form a particular class of generalized renewal
sequences, those for which all waiting times are exponential.

\vspace{.2cm}

We leave to the reader to deal with, i.e. make the appropriate conventions in
the cases when the sequence is finite because one of the waiting times is
infinite.

\vspace{.2cm}

\textit{Remark}\\
Though we formulated the original process as a Markov process on $\mathbf{G}$,
we could in fact have started from a generalized renewal process on $\mathbf{G}$
as described informally above (in that case, there would be no pressing reason
to singularize the first waiting time). All the arguments carry through,
yielding a generalized renewal process on $\mathbf{H}$. This looks even more
natural, because the continuous time Markov property is lost when going from
$\mathbf{G}$ to $\mathbf{H}$ by drag and drop, but the renewal property is
preserved.

\vspace{.2cm}

We have kept the discussion quite general in this subsection, but for the
application to affinities and cycles, we shall restrict to the case when
$\mathbf{H}$ is a micro-reversible cycle, i.e. a graph with $\geq 3$
configurations arranged in cyclic order.

\subsection{Conditioning for Markov processes} \label{ssec:cmp}

This section is about a topic that is likely to be very classical, but we have
not found a standard reference. The analogous computations for Markov chains are
totally elementary because they rely solely on conditional probabilities. One
could use a limiting argument to get, at least heuristically, the result for
Markov processes from the one for Markov chains. The direct computations in
continuous time require the use of conditional expectations\footnote{On
  conditional expectations, see the general references given at the beginning of
  Appendix \ref{ShortRemindMC}.}. They are lengthy and slightly more delicate,
so we shall only give the results. Let us note that there are some intimate
relations between conditioning and so-called taboo probabilities and taboo Green
functions, see e.g. \cite{jiang-qian-qian-2004}.

Suppose $(\mathbf{C},\Trans)$ is a Markov process with associated Markov matrix
$\M$ and graph $\mathbf{G}$. As usual, let $\Prob$ denote the law of this
process. Take a 
subgraph $\mathbf{H}$ of $\mathbf{G}$. Fix $T >0$. We want to describe the law
of the process conditioned to move along $\mathbf{H}$, i.e. such that all
jumps between time $0$ and $T$ are along edges of $\mathbf{H}$. Let
$\Prob_{\mathbf{H},T}$ be the probability law for the conditioned process.

We recall that it is in principle easy to collect samples of
$\Prob_{\mathbf{H},T}$. One simply collects samples of the original process
(with law $\Prob$) between $0$ and $T$
and keeps only those samples that obey the constraint that all jumps between
time $0$ and time $T$ are along edges of $\mathbf{H}$. 

The crucial ingredient is $\Gamma_{\mathbf{H},T}(\C,\C')$ defined as the
probability under the law $\Prob$ that a trajectory started at $\C$ at time $0$
ends at $\C'$ at time $T$ and jumps only along edges of $\mathbf{H}$ in between.
This is obviously $0$ when either $\C$ or $\C'$ is not a vertex of $\mathbf{H}$.

\vspace{.2cm}

\textit{Claim}\\
Using standard manipulations of conditional expectations, the Markov property
and homogeneity ($\Trans$ is time-independent) of the original process, one can
prove the following: for $0\leq s \leq t \leq T$, the
$\Prob_{\mathbf{H},T}$-probability that $\C_t=\C'$ when the process has been
observed up to time $s$ is given by
\[\Prob_{\mathbf{H},T}(\C_t=\C', \text{ knowing the past up to time }
s)=\frac{\sum_{\C''} \Gamma_{\mathbf{H},t-s}(\C_s,\C')
  \Gamma_{\mathbf{H},T-t}(\C',\C'')} {\sum_{\C''}
  \Gamma_{\mathbf{H},T-s}(\C_s,\C'')}.\] 

\vspace{.2cm}

Though the proof is not difficult, there are a number of steps and it would take
us too far from our main interest.  We advise the interested reader to build a
proof in the context of Markov chains, instead of Markov processes.

The above formula leads to two comments:\\
-- The right-hand side depends on the past only via $C_s$, meaning that the
conditioned process is still Markov.\\
-- Time $T$ appears explicitly on the right-hand side. This may be surprising
at first sight but a moment of thinking shows that this has to be so: even if we
look only at conditioned trajectories up to time $t$, their law depends on how
long the process is conditioned to move along $\mathbf{H}$ after $t$.

If a trajectory has moved along $\mathbf{H}$ between $0$ and $t$, one computes
easily the probability that between $t$ and $t+dt$ it will either stay where it
is or jump along $\mathbf{H}$. Defining a matrix $\N_{\mathbf{H}}$ by
$(\C' \vert \N_{\mathbf{H}}\vert \C)=0$ if either $\C$ or $\C'$ isn't a vertex
of $\mathbf{H}$, $(\C \vert \N_{\mathbf{H}}\vert \C)=(\C \vert \M \vert \C)$ if
$\C$ is a vertex of $\mathbf{H}$ and $(\C' \vert \N_{\mathbf{H}}\vert \C)=(\C'
\vert \M \vert \C)$ if $(\C,\C')$ is an edge of $\mathbf{H}$, we get
\[\frac{d}{dt}  \Gamma_{\mathbf{H},t}(\C,\C') = \sum_{\C''} (\C' \vert \N_{\mathbf{H}}\vert \C'')\Gamma_{\mathbf{H},t}(\C,\C''),
\] 
with initial condition $\Gamma_{\mathbf{H},0}(\C,\C')=1$ if $\C=\C'$ is a vertex
of $\mathbf{H}$ and $0$ otherwise.

Restricting $\Gamma_{\mathbf{H},t}$ and $\N_{\mathbf{H}}$ to vertices of
$\mathbf{H}$, we have the formula
\[\Gamma_{\mathbf{H},t}(\C,\C')=(\C' \vert e^{t\N_{\mathbf{H}}}\vert \C).\]

Thus we have:
\[\Prob_{\mathbf{H},T}(\C_t=\C', \text{ knowing the past up to time }
s)=\frac{\sum_{\C''} (\C'' \vert e^{(T-t)\N_{\mathbf{H}}}\vert \C')(\C' \vert
  e^{(t-s)\N_{\mathbf{H}}}\vert \C_s)}{\sum_{\C''} (\C'' \vert
    e^{(T-s)\N_{\mathbf{H}}}\vert \C_s)}.\]

  By the probabilistic interpretation, it is clear that all eigenvalues of
  $\N_{\mathbf{H}}$ have real part $\leq 0$. Under appropriate conditions,
  Perron-Frobenius theory will provide the existence of a unique eigenvalue
  with maximal real part, say $\lambda$ which will be real and correspond to
  left and right eigenvectors $(\eta \vert$ and $\vert \mu )$ with strictly
  positive components. In that case $e^{T\N_{\mathbf{H}}}\sim e^{\lambda T}\vert
  \mu )(\eta \vert$ for $T\to +\infty$, and
\[\lim_{T \to  +\infty} \Prob_{\mathbf{H},T}(\C_t=\C', \text{ knowing the past
  up to time } s)=e^{-\lambda (t-s)} (\eta \vert \C') (\C'\vert
e^{(t-s)\N_{\mathbf{H}}}\vert \C_s) (\eta \vert \C_s)^{-1}.\] This means
precisely that conditioning the Markov process $(\mathbf{C},\Trans)$ to move
(forever) along the subgraph $\mathbf{H}$ leads to a Markov process with
transition Matrix $\Trans_{\mathbf{H}}^{\text{cond}}$ (indexed by vertices of
$\mathbf{H}$) such that \be \label{eq:markcond} (\C' \vert
\Trans_{\mathbf{H}}^{\text{cond}}\vert \C)=(\eta \vert \C') (\C' \vert \Trans
\vert \C) (\eta \vert \C)^{-1} \text{ if } (\C,\C') \text{ is an edge of }
\mathbf{H} \text{ and } 0 \text{ otherwise}.  \ee Indeed, it is easy to check
that the Markov matrix $\M_{\mathbf{H}}^{\text{cond}}$ associated to
$\Trans_{\mathbf{H}}^{\text{cond}}$ is such that $
(\C'\vert e^{t\M_{\mathbf{H}}^{\text{cond}}}\vert \C)=e^{-\lambda t} (\eta \vert
\C') (\C'\vert e^{t\N_{\mathbf{H}}}\vert \C) (\eta \vert \C)^{-1}$.

Note that Perron-Frobenius theory applies if $\mathbf{H}$ is connected. We do
not prove this but observe that if $\mathbf{H}$ is connected an argument similar
to the one given at the end of Appendix \ref{ShortRemindMP} for $e^{t\M}$ for Markov processes shows that
$e^{t\N_{\mathbf{H}}}$ has, for every $t>0$, strictly positive matrix elements
between vertices of $\mathbf{H}$.

If $\mathbf{H}$ splits as a disjoint union of connected components,
Perron-Frobenius theory will apply to each component separately and a limiting
conditioned process on $\mathbf{H}$ will exist for $T \to +\infty$: the main
difference is that $\lambda$ will vary from one connected component to another.
But the formula \eqref{eq:markcond} will survive for an appropriate collection
$(\eta \vert \C)$ made by gathering the dominant eigenvectors for each connected
component.

\subsection{A summary of affinity class invariances} \label{ssec:ac}

We give a (non-exhaustive) list of probabilistic constructions that preserve the
affinity class. Most relevant for us are ``drag and drop'' and conditioning.

\vspace{.2cm}

\noindent \textbf{From a Markov process to a Markov chain} 

As recalled in Appendix \ref{ShortRemindMP}, to every Markov process
$(\mathbf{C},\Trans)$ one can associate a Markov chain $(\mathbf{C},\Pdt)$ such
that $\M=(\Pdt-\Id)\D$ where $-\D$ is the diagonal part of $\M=\Trans-\D$. Then
$\Trans$ and $\Pdt$ define the same graph, and when defined (i.e. for edges)
\[ \frac{(\C'\vert
  \Trans\vert \C)} {(\C\vert \Trans\vert \C')}=\frac{(\C'\vert
  \Pdt \vert \C)(\C\vert \D\vert \C)} {(\C\vert \Pdt \vert \C')(\C'\vert \D\vert
  \C')}.\] Taking the logarithm on both sides shows that affinity and its
discrete analog \eqref{eq:discr-aff} define the same class. 

\vspace{.2cm}

Our next aim is the behavior of the affinity class on subgraphs.  We start from
a Markov process $(\mathbf{C},\Trans)$ with associated graph $\mathbf{G}$, and
take a subgraph $\mathbf{H}$ of $\mathbf{G}$, both assumed to be
micro-reversible.

\vspace{.2cm}

\noindent \textbf{Restriction}

This construction is trivial, but has no particular probabilistic meaning: let
$\Trans_{\mathbf{H}}^{\text{rest}}$ be the restriction of the transition matrix to
$\mathbf{H}$. So $\Trans_{\mathbf{H}}^{\text{rest}}$ is a matrix indexed by
the vertices of $\mathbf{H}$ and
\[(\C' \vert \Trans_{\mathbf{H}}^{\text{rest}}\vert \C)=(\C' \vert \Trans \vert
\C) \text{ if } (\C,\C') \text{ is an edge of } \mathbf{H} \text{ and } 0 \text{
  otherwise}.\] From $\Trans_{\mathbf{H}}^{\text{rest}}$ we can define the exchange
$\mathbb{S}_{\mathbf{H}}^{\text{rest}}$, which is the restriction of
$\mathbb{S}$ to edges of $\mathbf{H}$. Trivially, the integral of
$\mathbb{S}_{\mathbf{H}}^{\text{rest}}$ or $\mathbb{S}$ along any walk on
$\mathbf{H}$ gives the same result.

\vspace{.2cm}

\noindent \textbf{Invariance under ``drag and drop''}

The intuitive definition of ``drag and drop'' (defined formally in Appendix
\ref{ssec:dd}, which should be consulted for reference) is the following. At
$t=0$ put a box at some vertex of $\mathbf{H}$ and start a walker on
$\mathbf{G}$. The box remains at its vertex until the walker gets there. When
the walker and the box are at the same vertex and the next jump of the walker is
along an edge of $\mathbf{H}$, the walker drags the box along.  This goes on as
long as the jumps of the walker are along $\mathbf{H}$. But as soon as the
walker prepares for a jump along an edge in $\mathbf{G}$ but not $\mathbf{H}$ he
drops the box. Then the box remains immobile and waits for the next encounter
with the walker and so on. So by successive drags and drops, the box itself
describes a walk on $\mathbf{H}$. There is an analogy that the reader may find
useful between hitch-hiking and a variant of ``drag and drop''(see the end of
Subsection \ref{ssec:dd}).

As explained in detail in Section \ref{ssec:ddmp}, under the procedure of ``drag
and drop'', if the walker samples a Markov process on $\mathbf{G}$, the
trajectory of the box samples a generalized renewal process on $\mathbf{H}$ and
in particular a Markov chain on $\mathbf{H}$. The corresponding stochastic
matrix $\Pdt_{\mathbf{H}}^{\text{dd}}$ is the Markov chain associated to the
Markov process associated to the restriction $\Trans_{\mathbf{H}}^{\text{rest}}$
of $\Trans$ to $\mathbf{H}$. Restriction preserves edge affinities on
$\mathbf{H}$, and going from a Markov process to its Markov chain may change edge
affinities, but preserves the affinity class.

So the cycle affinities are left invariant under ``drag and drop''.

\vspace{.2cm}
\clearpage
\noindent \textbf{Invariance under conditioning}

As explained in detail in Appendix \ref{ssec:cmp}, if the Markov process
$(\mathbf{C},\Trans)$ is conditioned to have all its jumps along edges of
$\mathbf{H}$, one ends up with a new Markov process, the configuration space
being made of the vertices of $\mathbf{H}$ and the transition matrix
$\Trans_{\mathbf{H}}^{\text{cond}}$ being defined by \eqref{eq:markcond} which
we repeat here for convenience: $\Trans_{\mathbf{H}}^{\text{cond}}$ is a matrix
indexed by the vertices of $\mathbf{H}$
\[(\C' \vert \Trans_{\mathbf{H}}^{\text{cond}}\vert \C)=(\eta \vert \C') (\C'
\vert \Trans \vert \C) (\eta \vert \C)^{-1} \text{ if } (\C,\C') \text{ is an
  edge of } \mathbf{H} \text{ and } 0 \text{ otherwise},\] where $(\eta \vert \C)$ is
a strictly positive function on the set of vertices of $\mathbf{H}$.  It is
plain that $\Trans_{\mathbf{H}}^{\text{rest}}$ and
$\Trans_{\mathbf{H}}^{\text{cond}}$ define the same affinity class.

So the cycle affinities are left invariant under conditioning. Let us recall
that to collect sample of the conditional probability one simply collects
samples of the original process (with law $\Prob$) and keeps only those samples
that obey the constraint that all jumps are along edges of $\mathbf{H}$.

\subsection{Discussion} \label{ssec:disc}

Let us pause for a moment to say a few words about the respective merits of
conditioning and ``drag and drop''. Conditioning seems at first sight to be more
natural, as it requires no extraneous structure (drag and drop introduces
another object, the box) and is defined for general micro-reversible
$\mathbf{H}$. However, conditioning is meant above as holding at all times (from
$0$ to $+\infty$): it is obtained via a limiting procedure from finite time
conditioning over $[0,T]$ (see the details in \ref{ssec:cmp}).  For finite time
conditioning, there are corrections to the time homogeneous description.  As $T
\to +\infty$ these corrections are generically exponentially small, but the
measure of the trajectories staying on $\mathbf{H}$ is exponentially small as
well. So the number of samples of the unconditioned system from which one
selects the ones fulfilling the constraint must be exponentially large. Hence
the perspectives of measuring the conditioned quantities are unclear (unless
there is an unbiased physical way to implement the constraint instead of simply
discarding the samples which do not fulfill them). On the other hand, ``drag and
drop'' needs to identify physically an extra structure (the box), and the ``drag
and drop'' motion is slower than the original motion, but does not suffer of the
exponential penalty of conditioning. However, the ``drag and drop'' motion is
described by a renewal process only when the vertices of $\mathbf{H}$ have
certain recurrence properties for the original Markov process trajectories. We
have mentioned in Section \ref{ssec:dd} that ``drag and drop'' possibly occurs
naturally in biological systems, and we could imagine many other situations, in
work-flows for instance. But we must admit that we have no natural
interpretation in the case of general non-equilibrium systems. However, one may
hope that for certain systems in which the dynamics is related to exchanges with
specific reservoirs, these exchanges give a clue to identify one or several
objects for which ``drag and drop'' is naturally observable.

We do not pursue this route here, and simply view conditioning and ``drag and
drop'' as important examples spelled in detail, showing that the affinity class,
and the corresponding cycle affinities, are remarkably robust observables that
remain invariant under a number of natural probabilistic constructions. These two
examples allow to observe processes or chains on subgraphs in terms of processes or
chains on graphs. Among all subgraphs, the simplest, most basic but somehow most
interesting for non equilibrium physics, subgraphs are cycles. In this context,
we call random motion on a cycle a ``noria''. Indeed, when a particle moves
along a cycle, one may keep track of its position, but also of its winding
number, i.e. one can count how many times the cycle has been described in a
given time interval, i.e. how many turns the noria has made. We shall see that
(cycle) affinity is the crucial quantity that describes the efficiency of the
noria, i.e. how fast the turns are made in average, in which direction, and what
is the size of fluctuations. To insist on the existence of fluctuations, we
might call it more precisely a mesoscopic noria. We now turn
to their detailed study, i.e. assume that the subgraph $\mathbf{H}$ is a cycle.

\section{A probabilistic interpretation of cycle affinities in norias}
\label{sec:pican}

In this section, we shall use repeatedly, and often without even mentioning it,
that the strong Markov property (for a precise definition, see the general
references at the beginning of Appendix \ref{ShortRemindMC}) holds for finite
state Markov chains and finite state Markov processes. Let us just describe
informally what this statement means.

The Markov property says that knowing the present (i.e. the configuration at a
given time $t$, $t=0,1,\cdots$ for chains, $t\in [0,+\infty[$ for processes) the
trajectory before $t$ and after $t$ are independent.

But what if $t$ is replaced by a random time $T$ ? For example, fix a
configuration $\C$ and let $T$ be the first time the trajectory visits $\C$,
i.e. $T$ is the smallest $s$ such that $\C_s=\C$. The strong Markov property
states that $\C_{T+t}$, ($t=0,1,\cdots$ for chains, $t\in [0,+\infty[$ for
processes) has exactly the same law as the original process started at $\C$,
i.e. with $\Pin(\C_0)=\delta_{\C_0,\C}$.

More generally, for a strong Markov process, this identity in law holds whenever
$T$ is a stopping time (again, for a precise definition, see the general
references at the beginning of Appendix \ref{ShortRemindMC}). Intuitively, a
stopping time $T$ is a random time for which one can decide if $T \leq s$ by
looking at the trajectory $\C_t$, $t\in [0,s]$, i.e. without knowledge of the
future of $s$. A deterministic time is a stopping time, so the strong Markov
property is in principle stronger than the Markov property. 

It is a theorem (and not a triviality) that finite state Markov chains and
finite state Markov processes do have the strong Markov property.

\subsection{Inhomogeneous random walks on cycles} \label{sec:irwc}

The general setting we need to study cycle affinities is a system with $\longc \geq
3$ configurations labeled $1,\cdots,\longc$ (configuration $m$ and $m+\longc$ are
identified in all subsequent formul\ae). We choose arbitrarily an orientation
of the cycle: a jump from $m$ to $m+1$ (resp. $m-1$) is said to be clockwise
(resp. anti-clockwise). We view the process as the motion of a particle along
the cycle, jumping from time to time from a site to one of its two neighbors.
As established in the previous subsection, whatever the initial Markov process
was, as far as the cycle $\mathbf{H}$ in the initial graph is concerned, the
rules are of the following kind.

At $t=0$ the particle starts at some $m$ on the cycle. We make the convention
that the first waiting time does not play a special role. This is automatic for
conditioning, but not for drag and drop (in that case, either we start the
walker at $m$, or start time at the first passage of the walker at $m$). The
particle waits at $m$ for a (random) time $T$ (whose law depends only on $m$)
and then jumps to $m'=m\pm 1$ with probability $p_{m}^{\pm}$. It waits there for
a random time $T'$ (whose law depends on the past only via the position $m'$)
and at $T+T'$ jumps to $m''=m'\pm 1$ with probability $p_{m'}^{\pm}$. It waits
there for a random time $T''$ (whose law depends on the past only via the
position $m''$) and so on.

In particular the successive positions $m,m',m'',\cdots$ form a Markov chain with
a stochastic matrix whose only non-vanishing elements are $(m\pm 1 \vert \Pdt
\vert m)=p_{m}^{\pm}$ (use the periodic boundary conditions when necessary) with
$p_{m}^{\pm} > 0$ and $p_{m}^{+}+p_{m}^{-}=1$ for $m=1,\cdots,\longc$. The law of the
random time spent at $m$ could in principle be computed for each $m$ from the
knowledge of the original Markov process, a rather simple task for
conditioning (where all waiting times are exponential and with easily
computable parameters), but a very involved one for ``drag and drop''. 

\subsubsection{Winding numbers} \label{sssec:wn}

We define a sequence $W_k$ as follows: $\longc W_k$ is the number of clockwise jumps
minus the number of anti-clockwise jumps among the $k$ first jumps. The process
$(\longc W_k)_{k=0,1,\cdots}$ is an inhomogeneous random walk on the integers, the
inhomegeneity being periodic in space with period $\longc$.  If the initial position
was $m$, $m+\longc W_k$ (taken modulo $\longc$) is the position just after the
$k^{\text{th}}$ jump. But $W_k$ carries more information: by construction, $W_k$
changes by $-1$, $0$ or $1$ between two successive passages at $m$ for
$m=1,\cdots,\longc$, so it keeps a memory of the number of turns. This is the reason
why we call $W_k$ the winding number up to (and including) the $k^{\text{th}}$
jump.

There are two natural ways to measure time: time in the original Markov process,
$t$, and time as number of steps in the cycle Markov chain, $k$. If $t$ belongs
to the interval between jump $k$ and jump $k+1$, so that the winding number is
$W_k$, we set $\tilde{W}_t \equiv W_k$. Observe that $\tilde{W}_t$, $t\in [0,
+\infty[$ and by extension $W_k$, $k=0,1,2,\cdots$ are exchange processes as
defined in \ref{ssec:ep}.

The first question we shall ask is the probability that the winding number will
ever reach a certain value. This question does not involve a time
parameterisation, and has a simple answer. 

The second question will be about the distribution of the time it takes to reach
a certain winding number, and this question, closely related to the efficiency
of the noria, depends on the time
parameterisation. The answer is more complicated and less explicit, but exhibits
some simple general symmetries, strongly reminiscent of out-of equilibrium relations.

\vspace{.2cm}

The quantity $p_m^+/p_{m+1}^-$ quantifies the relative tendency to
traverse edge $(m,m+1)$ in one direction or another. In the rest of this
section, we set 
\[ e^A\equiv \prod_{m=1}^\longc \frac{p_m^+}{p_m^-} \] and see directly that $A$ is
the affinity along the cycle oriented clockwise. Thereafter we use the simple
name ``affinity'' for the quantity $A$.

\subsubsection{Algebraic preliminary}  \label{subsubsec:ap}

We start with an algebraic preliminary which is the crucial ingredient for the
following elaborations. 

We suppose given a collection of indeterminates $x_m$, $m=1,\cdots,\longc$. We look
for the solution(s) of the system 
\be \label{eq:gensys-}
f_m^-=x_m(p_m^-+p_m^+f_{m+1}^-f_m^-), \ee 
with periodic boundary conditions, for the unknown $f_m^-$, $m=1,\cdots,\longc$, and for  the solution(s) of the system 
\be \label{eq:gensys+}
f_m^+=x_m(p_m^++p_m^-f_{m-1}^+f_m^+), \ee
with periodic boundary conditions, for the unknown $f_m^+$, $m=1,\cdots,\longc$.

Later, we shall give a probabilistic interpretation of these systems, where the
indeterminates $x_m$ will be real numbers in $]0,1]$, or complex numbers in the
closed unit disc. We set $F^-\equiv \prod_{m=1}^\longc f_m^-$ and $F^+\equiv
\prod_{m=1}^\longc f_m^+$.

\vspace{.2cm}

\textit{Claim}\\
Each system \eqref{eq:gensys-} or \eqref{eq:gensys+} has generically two
solutions. In particular, $F^+$ as well as $F^-$ can take two values. Moreover,
either $F^-F^+=1$ or $F^-=F^+ e^{-A}$.

\vspace{.2cm}

\textit{Proof}\\
We start with system \eqref{eq:gensys-}. This system is a discrete
analog of a Ricatti differential equation, and it can be linearized by the
standard trick. Let $g_0$ be arbitrary and define iteratively $g_m\equiv
f_m^-g_{m-1}$ for 
$m=1,\cdots,\longc,\longc+1$. Observe that $g_\longc=F^-g_0$ and $g_{\longc +1}=F^-g_1$ : the
sequence $f_m$ is periodic by construction but the sequence $g_m$ is not, and
$F^-$ appears as an holonomy.  Then \eqref{eq:gensys-} turns into 
\be
 \label{eq:gm} 
g_m=x_m(p_m^-g_{m-1}+p_m^+g_{m+1}) \text{ for }  m=1,\cdots,\longc.
\ee
This two-terms recursion relation can be transformed in a vector one-term
recursion relation:
\[ \left(\begin{smallmatrix} g_{m+1} \\ g_{m}\end{smallmatrix} \right)=
\mathbb{F}_m \left(\begin{smallmatrix} g_{m} \\ g_{m-1}\end{smallmatrix} \right)
\]
for $m=1,\cdots,\longc$, where
\be \label{eq:matFm} \mathbb{F}_m\equiv\left(\begin{smallmatrix} 1/(x_mp_m^+)& -p_m^-/p_m^+ \\ 1 &
    0\end{smallmatrix}\right). \ee
We set
\[ \mathbb{F}=\left(\begin{smallmatrix}\mathbb{F}_{11} & \mathbb{F}_{12} \\
    \mathbb{F}_{21} & \mathbb{F}_{22} \end{smallmatrix}\right) \equiv\mathbb{F}_\longc \cdots
\mathbb{F}_1.\]
One finds by iterating the above formula that 
\be
 \label{eq:Fn} F^-\left(\begin{smallmatrix} g_{1} \\ g_{0}\end{smallmatrix} \right)
=\left(\begin{smallmatrix} g_{\longc +1} \\ g_{\longc }\end{smallmatrix}
\right)=\mathbb{F} \left(\begin{smallmatrix} g_{1} \\ g_{0}\end{smallmatrix}
\right)
\ee 
so that $F^-$ is an eigenvalue of the transfer (or Bloch-Floquet or $\cdots$
depending on the community) matrix $\mathbb{F}$, i.e. a solution of 
\be
 \label{eq:F-}
 (F^-)^2-F^- \, \text{Tr} \, \mathbb{F}+\text{Det} \, \mathbb{F}=0.  
\ee 
This formula shows clearly that $F^-$ is a cyclic invariant, because $\text{Tr}
\, \mathbb{F}$ and $\text{Det} \, \mathbb{F}$ are. The trace $\text{Tr} \,
\mathbb{F}$ has a complicated expression in general, but the determinant
$\text{Det} \, \mathbb{F}=\prod_{m=1}^\longc \text{Det} \, \mathbb{F_m}
=\prod_{m=1}^\longc p_m^-/p_m^+ =e^{-A}$, a simple function of the cycle affinity.

Eq.\eqref{eq:F-} is a quadratic equation, which has generically two solutions.
If one is chosen, the ratio of the components of the corresponding eigenvector
$g_{1}/g_{0}=f_1^-$ is fixed, namely
$f_1^-=(F^- -\mathbb{F}_{11})/\mathbb{F}_{21}$, and then all other
$f_{\cdot}^-$'s as well using \eqref{eq:gensys-}.

To summarize, we have shown that the system \eqref{eq:gensys-} has generically
exactly two solutions. 

The system \eqref{eq:gensys+} can be solved in an analogous way. It is easily
seen that if one takes an arbitrary $g_{\longc +1}$ and defines iteratively $g_m\equiv
f_m^+g_{m+1}$ for $m=\longc,\cdots,1,0$ the $g_{\cdot}$'s do again satisfy
\eqref{eq:gm} (though they may differ from the $g_{\cdot}$'s introduced using
the $f_{\cdot}^-$'s). This time $g_0=F^+g_\longc$ and $g_{1}=F^+g_{\longc +1}$ and we infer
that \be
 \label{eq:F+}
(F^+)^{-2}-(F^+)^{-1} \, \text{Tr} \, \mathbb{F}+\text{Det} \, \mathbb{F}=0.
\ee

So $F^-$ and $1/F^+$ are roots of the same quadratic equation, and there are two
possibilities: either $F^-F^+=1$ or $F^-=F^+\text{Det} \, \mathbb{F}$. As
$\text{Det} \, \mathbb{F}=e^{-A}$ the proof is completed. \hfill $\square$

\vspace{.2cm}

We shall see in the sequel (see Sections \ref{subsubsec:wtmcs} and \ref{subsubsec:wtgc}) that
for our probabilistic aims, the second possibility, namely  $F^-=F^+\text{Det} \,
\mathbb{F}$, is the relevant one.  Then an explicit computation yields 
\be
\label{eq:ratiopm1} \frac{f_1^-}{f_0^+}=-\frac{
  \mathbb{F}_{12}}{\mathbb{F}_{21}}.
\ee
The ratios $f_m^-/f_{m-1}^+$ are given by analogous formul\ae\ obtained by
replacing the product $\mathbb{F}\equiv\mathbb{F}_\longc \cdots \mathbb{F}_1$ by the
appropriate circular permutation. Apart from the fact that these ratios involve
no square root and are plain rational functions of all the arguments, the
explicit expressions are complicated in general.

\vspace{.2cm}

We return now to our primary interest, the cyclic Markov chain. The strategy in
the following sections is always the same: use the (strong) Markov property to
obtain recursion relations of the type studied above among the quantities of
interest, then use the periodicity along the cycle to solve the recursion
relations.

\subsubsection{Clockwise  and anti-clockwise cycles}  \label{subsubsec:cacc}

The easiest question to answer is very classical and its solution can be found
in textbooks (though the proof we give is not totally standard): will the winding
number will ever reach $\pm 1$ ?

\vspace{.2cm}

\textit{Claim}\\
If $A\geq 0$ (resp. $\leq 0$) the probability to reach
winding number $1$ is $1$ (resp. $e^A$) and the probability to reach winding
number $-1$ is $e^{-A}$ (resp. $1$). 

\vspace{.2cm}

The complete proof has to wait until \ref{subsubsec:wtmcs}, but for orientation
we give a very simple argument leading to a slightly weaker result. The (strong)
Markov property is the crucial ingredient.

\vspace{.2cm}

\textit{Sketch of proof}\\
Let
$\pi_m^-$ be the probability that the winding number of a trajectory started at
$m$ ever reaches the value $-1/\longc$. Then by the Markov property 
\be
\label{relrecurrpi-} 
\pi_m^-=p_{m}^{-} + p_{m}^{+} \pi_{m+1}^-\pi_m^-.  
\ee 

The meaning of this equation is clear : either the first jump is anti-clockwise
(probability $=p_{m}^{-}$) and the winding number reaches its target $-1/\longc$ or
the first jump is clockwise (probability $=p_{m}^{+}$), and then the particle
has ``lost'' a winding $1/\longc$ so it has to go from $m+1$ to $m$ with winding
number $-1/\longc$ to compensate (probability $\pi_{m+1}^-$), and take a new chance.

Notice that \eqref{relrecurrpi-} is simply \eqref{eq:gensys-} with all $x_m$'s
replaced by $1$. In this simple special case, an ad hoc argument works because
$f_m^-=1$ for $m=1,\cdots,\longc$ is obviously a solution. The probability that
starting from $m$ the winding number reaches $-1$ is, by the Markov property
again, $\prod_{l=0}^{\longc-1}\pi_{m-l}^-$, which is independent of $m$. So we denote
this probability simply by $\Pi^-\equiv \prod_{l=0}^{\longc-1}\pi_{m-l}^-$. We
rewrite \eqref{relrecurrpi-} as \be
\label{eq:prodwind} p_m^-(1-\pi_m^-) =p_m^+ \pi_m^- (1-\pi_{m+1}^-).  \ee A
first consequence is that if $\pi_m^- =1$ for some $m$ then also $\pi_{m+1}^-=1$
and so on, so that all $\pi_m^-$'s are equal to one, and the probability to
reach winding number $-1$ is unity. On the other hand, in the case when no
$\pi_m^-$ equals one, \be \Pi^-=\prod_{m=1}^\longc\frac{p_m^-}{p_m^+}  = e^{-A}. \ee
Indeed, in \eqref{eq:prodwind} take the product over all $m$'s in the cycle and
simplify both sides by $\prod_{m=1}^\longc (1-\pi_m^-) \neq 0$ to get $\prod_{m=1}^\longc
p_m^- =\prod_{m=1}^\longc p_m^+ \prod_{m=1}^\longc \pi_m^-$, i.e. $\Pi^-= \prod_{m=1}^\longc
p_m^-/p_m^+$. As $\Pi^-$ is a probability, this is possible only if
$\prod_{m=1}^\longc p_m^-/p_m^+ \leq 1$.

We have already proved the following: if $A \leq 0$ then $\Pi^- =1$, i.e.
winding number $-1$ is reached with probability $1$, and if $A \geq 0$ then
$\Pi^- \in \{1,e^{-A}\}$. We could reproduce the above argument with $\pi_m^+$,
the probability that the winding of a trajectory started at $m$ ever reaches the
value $1/\longc$, and $\Pi^+\equiv  \prod_{l=0}^{\longc-1}\pi_{m+l}^+$. We would get
: if $A \geq 0$ then $\Pi^+ =1$, i.e.  winding number $1$ is reached with
probability $1$ and if $A \leq 0$ then $\Pi^+ \in \{1,e^{A}\}$. In particular,
if $A=0$ then the probability to reach winding numbers $-1$ and $1$ is unity,
and by the (strong) Markov property, the probability to reach any winding number an
infinite number of times is also unity: the winding number $W_k$ will oscillate
and take arbitrarily large positive and negative values as $k\rightarrow
+\infty$. \hfill $\square$

\vspace{.2cm}

These arguments suggest that in fact \be
\label{eq:Pi-+} \Pi^- = \min \{1, e^{-A}\} \text{ and } \Pi^+ = \min \{1, e^{A}
\}.\ee Though it is intuitive that if $A > 0$ there is a systematic drift
towards positive winding number, hence a finite probability to never reach
winding number $-1$, it takes a deeper argument to get a proof. This is our next
aim.

\subsubsection{First passage times: Markov chain case} \label{subsubsec:wtmcs} 

We generalize a bit the previous discussion. We let $F^-$ (resp. $F^+$) be the
generating function for the number of steps it takes to reach winding number
$-1$ (resp. $1$), the parameter being $z$. By definition, $F^-\equiv \sum_{k
  \geq 1} F_k^- z^k$ where $F_k^-$ is the probability that it takes exactly $k$
steps to reach winding number $-1$. In the same way $F^+\equiv \sum_{k
  \geq 1} F_k^+ z^k$ where $F_k^+$ is the probability that it takes exactly $k$
steps to reach winding number $1$. Note that  $F^\pm(z=1)=\Pi^\pm$.

Defining $f_m^-$ to be
the generating function for the number of steps it takes to reach winding number
$-1/\longc$ starting from $m$, the following holds.

\vspace{.2cm}

\textit{Claim}\\
The $f_m^-$'s solve the
system \eqref{eq:gensys-} for all $x_m$'s equal to $z$, namely \be \label{eq:gensys-xm=z}
f_m^-=z(p_m^-+p_m^+f_{m+1}^-f_m^-)\text{ for } m=1,\cdots,\longc \text{ with periodic boundary
conditions,}\ee and $F^-=\prod_{m=1}^\longc f_m^-$.

\vspace{.2cm}

\textit{Proof}\\
That the $f_m^-$'s are periodic modulo $M$ is included in their very definition.
Consider the right-hand side of \eqref{eq:gensys-xm=z}. The factor $z$ is the
weight of the first step. If the first step is anti-clockwise (probability
$p_m^-$) winding number $-1/\longc$ is reached after one step. Otherwise (probability
$p_m^+$) the first step takes to winding number $1/\longc$ and then the number of
steps to reach winding number $-1/\longc$ is, by the (strong) Markov property, the
sum of two independent contributions, one with generating function $f_{m+1}^-$
and one with generating function $f_{m}^-$. Finally, $F^-=\prod_{m=1}^\longc
f_m^-$, because $F^-$ is the generating function for a sum of $\longc$ independent
(the strong Markov property again) random variables with generating functions
$f_m^-$, $m=1,\cdots,\longc$. \hfill $\square$

\vspace{.2cm}

Of course an analogous results holds for $F^+$.  For the rest of this discussion
we set $x_m=z$ for all $m$'s in the formul\ae\ from \ref{subsubsec:ap}.

\vspace{.2cm}

\underline{\textit{Claim of our main result (Markov chain case)}}\\
The generating functions $F^-$ and $F^+$ satisfy $F^-=F^+ e^{-A}$,
  and in particular \eqref{eq:Pi-+} holds.

\vspace{.2cm}

\textit{Proof}\\
The definition of $F^-$ as a normally convergent sum in the closed unit disc
$|z| \leq 1$ shows immediately that $F^-$ is in fact an analytic function in the
open unit disc $|z| < 1$, continuous up to the boundary i.e. in the closed unit
disc $|z| \leq 1$ and the same holds for each $F_m^-$. Moreover $F^-(z=1)=\Pi^-$, while by definition $f_m^-(z)\sim p_m^- z$ at small $z$.

As $F^-$ is real for $z \in ]0,1]$, obviously the discriminant of the quadratic
equation \eqref{eq:F-} has to be $\geq 0$ in this parameter range.

Now, at small $z$, the upper left corner is the dominant term in each
$\mathbb{F}_m$ (defined in \eqref{eq:matFm} where $z$ is substituted for $x_m$) so that $\text{Tr} \, \mathbb{F}\sim z^{-\longc}/(\prod_{m=1}^\longc
p_m^+)$ at small $z$. Taking $z$ small and positive, we see that to prevent
explosion the branch in the solution of \eqref{eq:F-} has to be \be
\label{eq:FF-} F^-=\frac{\text{Tr} \, \mathbb{F} - \sqrt{(\text{Tr} \,
    \mathbb{F})^2-4\text{Det} \, \mathbb{F}}}{2}, \ee where the determination of
the square-root is the analytic continuation of the positive square root at
small $z$.

The generating function $F^+$ for the number of steps it takes to reach winding
number $1$, the parameter being $z$, is related to \eqref{eq:gensys+} in the
same way. Now $F^+$ is small at small $z$, so that the relevant solution of
\eqref{eq:F+} at small $z$ is \be \label{eq:FF+} \frac{1}{F^+}=\frac{\text{Tr}
  \, \mathbb{F} + \sqrt{(\text{Tr} \, \mathbb{F})^2-4\text{Det} \,
    \mathbb{F}}}{2}.  \ee where the determination of the square-root is the
analytic continuation of the positive square root at small $z$. In particular we
find that
\[ \frac{F^-}{F^+}=\text{Det} \, \mathbb{F} =e^{-A}\]
is valid everywhere by analytic continuation. As a consequence, taking $z
\rightarrow 1^-$ one finds
\[ \frac{\Pi^-}{\Pi^+}=\text{Det} \, \mathbb{F} = e^{-A} ,\] which together with
the results in \ref{subsubsec:cacc} proves \eqref{eq:Pi-+}. \hfill $\square$

\vspace{.2cm}

A few remarks are in order:\\
-- The proof that we have given for the identity $F^-=F^+ e^{-A}$ is
analytic (via generating functions and singularity analysis), but it would
deserve a \textit{good} combinatorial proof. It is clear that if we compared the
generating functions for winding number $\pm 1$ (without the restriction that it
is the first passage) we would obtain a one-to-one correspondence preserving
length (i.e. the number of steps) by simply reversing paths, and the ratio would obviously be
$\prod_{m=1}^\longc p_m^-/p_m^+$. In the same way, if we compared generating
functions for winding number $-1/\longc$ starting at $m$ and for winding number $1/\longc$
starting at $m-1$ (without the restriction that it is the first passage) the
ratio would be $p_m^-/p_{m-1}^+$. Whereas the ratio remains simple (and the
same) in the first case (full cycle) when we restrict to first passages, the
ratio becomes more involved in the second case (piece of cycle) when we restrict
to first passages: by \eqref{eq:ratiopm1}, the ratios $f_m^-/f_{m-1}^+$ are
given by explicit but tedious formul\ae, with one salient feature however, they
are rational functions of the parameters of the Markov chain.     \\
-- The proof we have given that if $\prod_{m=1}^\longc p_m^-/p_m^+ <1$ there
is finite probability to never reach winding number $-1$ is also of analytic
nature, but would deserve a purely probabilistic proof. This could be obtained
either as a consequence of the individual ergodic theorem or of the renewal
theorem, or of other refinements of the strong law of large numbers.

\subsubsection{First passage times: general case} \label{subsubsec:wtgc} 

In the beginning of \ref{sec:irwc} we have explained why concentrating on a
cycle for general finite state Markov processes leads to a Markov chain on the
cycle and to waiting times between the jumps which depend only on the position
on the cycle but are in general not exponential. 

So take $x_m\equiv x_m(\lambda)$ as the Laplace transform of the law of the time
during which the particle remains at $m$ before jumping. If the probability that
this time belongs to $[0,t[$ is $u_m(t)$ then $x_m(\lambda)\equiv
\int_0^{+\infty} e^{-\lambda t} \, du_m(t)$. As the $u_m$'s are derived (though
in a complicated way for ``drag and drop'') from the exponential waiting times
of a finite state Markov process, they are quite nice, i.e. continuous,
vanishing at $0$ and going to $1$ exponentially fast at large $t$ so that
$x_m(\lambda)$ has finite derivatives of all orders (giving the moments) at
$\lambda=0$ and $\lim_{\lambda \to +\infty}x_m(\lambda)=0$.

Let $T^-$ be the (random) time it takes to reach winding number $-1$ and $T^+$
be the (random) time it takes to reach winding $1$. We do not exclude \textit{a
  priori} that for some trajectories, the particle never reaches winding number
$-1$ or $+1$ and on such events we define $T^-$ or $T^+$ to be $+\infty$.

Let $f_m^-\equiv f_m^-(\lambda)$ (resp.  $F^-\equiv F^-(\lambda)$) stand for the
Laplace transform of the law of the time it takes to reach winding number $-1/\longc$
starting from $m$ (resp. to reach winding number $-1$), and let $f_m^+\equiv
f_m^+(\lambda)$ (resp.  $F^+\equiv F^+(\lambda)$) stand for the Laplace
transform of the law of the time it takes to reach winding number $1/\longc$ starting
from $m$ (resp. to reach winding number $1$). In equations, $F^-(\lambda)\equiv 
\int_0^{+\infty}e^{-\lambda t}\, dP(T^-<t)$ and $F^+(\lambda)\equiv 
\int_0^{+\infty} e^{-\lambda t}\, dP(T^+<t)$.
In these definitions, time stands
for the real time of the original Markov process, not for the number of steps.

\vspace{.2cm}

\underline{\textit{Claim of our main result (general case)}}\\
The functions $f_m^-(\lambda)$ (resp. $f_m^+(\lambda)$) solve the system
\eqref{eq:gensys-} (resp. \eqref{eq:gensys+}) with data $x_m(\lambda)$. The
functions $F^-(\lambda)$ and  $F^+(\lambda)$ are related by $F^-(\lambda) =
F^+(\lambda) e^{-A}$.

\vspace{.2cm}

\textit{Proof}\\
To show that the functions $f_m^-(\lambda)$ (resp. $f_m^+(\lambda)$) solve the
system \eqref{eq:gensys-} (resp. \eqref{eq:gensys+}) with data $x_m(\lambda)$,
the same probabilistic argument as before can be repeated word for word.

Because $\lim_{\lambda \to +\infty}x_m(\lambda)=0$, the same simple
argument as before allows to choose the appropriate branch for the quadratic
equation. Hence the following relation holds between the Laplace transforms
$F^{\pm}(\lambda)$ of the laws of the time it takes to reach winding
number $\pm1$: \be \label{eq:flucrel}
\frac{F^-(\lambda)}{F^+(\lambda)}=\text{Det} \, \mathbb{F}= e^{-A},  \ee so that
the cycle affinity still governs the relationship between the distribution of
first passage times at winding numbers $-1$ and $1$. The ratio is independent from the $x_m$'s, namely from the distributions of the waiting times at the various positions on the cycle.\hfill $\square$

\vspace{.2cm}

By \eqref{eq:ratiopm1}, the
ratios $f_m^-(\lambda)/f_{m-1}^+(\lambda)$ are more complicated. They 
are given by explicit but tedious formul\ae , with one salient feature however: they
are rational functions of the parameters of the Markov chain. 

Relation \eqref{eq:flucrel} can be viewed as a random time analog (or dual) of
the usual finite time out-of-equilibrium relations. The usual question would be
to ask whether there is a relation between the probabilities to see a winding
number $w$ versus $-w$ at a fixed time $t$: time is fixed, $\tilde{W}_t$ (as
defined in Section \ref{sssec:wn}) is random and one compares in some way
$p(\tilde{W}_t=w)$ and $p(\tilde{W}_t=-w)$.  But here, $T^+$ is the time to
reach winding number $1$, $T^-$ is the time to reach winding number $-1$ and
we compare the laws of $T^+$ and $T^-$.

To make it look even more familiar, we may define $T^{(w)}$ for an arbitrary
\textit{integer} $w$ as the time needed to
reach winding number $w$, and $F^{(w)}(\lambda)$ as the  Laplace
transform of the law of $T^{(w)}$, so that $F^{\pm}=F^{(\pm 1)}$. The renewal property entails that
$F^{(w)} $ is simply the $w^{\text{th}}$ (resp $(-w)^{\text{th}}$) power of $F^+$
(resp. $F^-$) for $w \geq 0$ (resp. $w \leq 0$). Then, for each integer $w$ 
\[ 
\frac{F^{(-w)}(\lambda)}{F^{(w)}(\lambda)}=e^{-wA}.
\]
Note that this is an exact relation valid for finite winding numbers and for any
cycle in a more general system of transitions as long as the procedure to define
the process of the cycle preserves the affinity class, as shown for instance for
``drag and drop'' and conditioning. However, the relation is established under
the assumption that the first waiting time plays no special role. This is
natural for conditioning but not for ``drag and drop'' for instance. Then
letting $\hat{F}^{(w)}(\lambda)$ denote the ``real'' quantity (vaguely speaking,
taking into account the time it takes to reach the cycle) the law of large
numbers leads to the fact that $[\hat{F}^{(w)}(\lambda)]^{1/w}$ behaves for large
$w\to +\infty$ (resp. $w\to -\infty$) as $F^+(\lambda)$ (resp. $F^-(\lambda)$)
so that relation \eqref{eq:flucrel} is observable as an asymptotic result in
general:
\[
\lim_{w\to \pm \infty}
\frac{[\hat{F}^{(-w)}(\lambda)]^{1/w}}{[\hat{F}^{(w)}(\lambda)]^{1/w}}= e^{-A}.
\]

\vspace{.2cm}

\textit{Remark}\\
There is another interpretation of the identity $F^-(\lambda) = F^+(\lambda)
e^{-A}$. Let $P^+$ (resp. $P^-$) be the law on $[0,+\infty]$ of the time $T^+$ (resp. $T^-$) it
takes to reach winding number $1$ (resp. $-1$), so that $F^+(\lambda)$ (resp.
$F^-(\lambda)$) is nothing but the Laplace transform of $P^+$ (resp. $P^-$).
Then $P^+$ is absolutely continuous with respect to $P^-$ and 
\[ P^-= e^{-A}P^+ + (1-e^{-A})\delta_{+\infty},\] i.e. the Radon-Nykodim
  derivative  
$dP^+/dP^-$ is equal to the constant  $e^{A}$.

\subsubsection{Mean first passage time and efficiency}

The previous arguments have lead to a general symmetry relation between
clockwise and anti-clockwise first passage times, but as stated above the
computation of the law of the individual clockwise or anticlockwise first
passage times is complicated.

We illustrate this first level of complexity by giving a formula for the mean
first passage time. This is the most straightforward measure of the efficiency
of the cycle, i.e. the average speed at which cycles are performed by the noria.

We assume here that $A >0$ so that $\Pi^+=1$, $\Pi^-=e^{-A}<1$ and the winding
number grows in average. This suggests strongly that the mean first passage
times at positive winding numbers are finite. We shall not give a proof of this
(this is standard and not really difficult, but would take us too far) but we
give the formula.

We let $l_m$ denote the mean waiting time at $m$ for $m=1,\cdots,\longc$ and $\tau_m^+$
be the mean first passage time at winding $1/\longc$ starting from $m$, so that, by
the strong Markov property, the
mean first passage time at winding $1$ is, whatever the starting point,
$\sum_{m=1}^\longc \tau_m^+\equiv \Esp{T^+}$. For an arbitrary generalized renewal process,
the finiteness of $l_m$ is not guaranteed and should be included as an hypothesis.
For the constructions we use in this work (conditioning or drag and drop on a
subgraph, starting from a finite state Markov process), it is automatic.

\vspace{.2cm}

\textit{Claim}\\
The equation \be \tau_m^+=p_m^+l_m+p_m^-(l_m+\tau_{m-1}^++\tau_m^+)
\label{eq:avetime} \ee holds. 

\vspace{.2cm}

\textit{Proof}\\
We give two arguments:\\
-- The first one is related to the generating functions in the parameter
$\lambda$ introduced in \ref{subsubsec:wtgc}, because average times are first
moments, so they can be computed by taking a derivative with respect to
$\lambda$ at $\lambda=0$. Thus $l_m$ (resp.  $\tau_m^+$) is nothing but
$-\frac{dx_m(\lambda)}{d\lambda}(\lambda=0)$
(resp.$-\frac{df_m^+(\lambda)}{d\lambda}(\lambda=0)$), so \eqref{eq:avetime} can
be obtained by taking the derivative of \eqref{eq:gensys+} with respect to
$\lambda$ at $\lambda=0$. Notice that $x_m(\lambda=0)=1$ is always true, but
$f_m^+(\lambda=0)=1$ holds only
because $\Pi^+=1$.\\
-- The second argument, a direct probabilistic derivation relying on the strong
Markov property, is even more
transparent: it takes in average $l_m$ units of time to make the first step; if
it is clockwise, (probability $=p_{m}^{+}$), we are done, but if the first step
is anti-clockwise (probability $=p_{m}^{-}$) reaching winding number $1/\longc$ takes
in average, on top of $l_m$, time $\tau_{m-1}^+$ to hit winding number $0$ and time
$\tau_m^+$ to hit winding number $1/\longc$. \hfill $\square$

\vspace{.2cm}

\textit{Claim}\\
The mean first passage time at winding number $1$, $\Esp{T^+}$, is given by
\be \label{eq:afpt} \Big(\prod_{m=1}^\longc p_m^+ - \prod_{m=1}^\longc p_m^-\Big)\Esp{T^+}=\sum_{m=1}^\longc \sum_{k=1}^\longc \Big(\prod_{1\leq i <k}p_{m+i}^+\Big) l_{m+k}
\Big(\prod_{k < j \leq \longc}p_{m+j}^-\Big).\ee

\vspace{.2cm}

This formula is still elegant, but already complicated. The other moments of the
first passage time $T^+$ can be computed by analogous tricks but require more
efforts, and the final expressions require more space.

\vspace{.2cm}

\textit{Proof}\\
Equation \eqref{eq:avetime} can be rewritten: 
\[\tau_m^+=\frac{l_m}{p_m^+}+\frac{p_m^-}{p_m^+}\tau_{m-1}^+.\] Using this relation
repeatedly, we get $\tau_\longc^+$ in terms of $\tau_{\longc -1}^+$, then in
terms of $\tau_{\longc -2}^+$
and so on, until we finally get a closed equation because $\tau_{0}^+=\tau_\longc^+$. This
leads to: 
\[\Big(\prod_{m=1}^\longc p_m^+ - \prod_{m=1}^\longc p_m^-\Big) \tau_\longc^+= 
\sum_{k=1}^\longc \Big(\prod_{1\leq i <k}p_i^+\Big) l_k \Big(\prod_{k < j \leq
  n}p_j^-\Big).\]
Note that $\prod_{m=1}^\longc p_m^+ - \prod_{m=1}^\longc p_m^- >0$, which is equivalent to
$A>O$, is our working
hypothesis, leading to growth of the winding number. Under the opposite
hypothesis, the formula would be inconsistent as it would lead to negative
mean first passage times.  The other $\tau_m^+$'s are easily obtained by a cyclic
permutation, finally leading to \eqref{eq:afpt}. \hfill $\square$

\vspace{.2cm}

The mean first passage time at winding number $w=0,1,2,\cdots$ is $w\Esp{T^+}$.
The strong law of large numbers will imply that the first passage time at
winding number $w$, which is a sum of $w$ independent random variables with mean
$\Esp{T^+}$ will differ from its mean $w \Esp{T^+}$ by an $o(w)$ with
probability going to $1$ as $ w\rightarrow +\infty$. By a standard trick this
results can be expressed in an equivalent way as: \be \label{eq:speed} \lim_{t
  \to +\infty} \tilde{W}_t/t = 1/\Esp{T^+} \text{ with probability }1.  \ee So
when the noria works for a long time fluctuations are suppressed and a
``classical'' behavior emerges : the noria turns at speed $1/\Esp{T^+} +o(1)$
and can thus be fully characterized by its efficiency.

\vspace{.2cm}

Formula \eqref{eq:speed} can be checked against a particular case: if we assume
that the cycle is described by a Markov process, i.e. that the waiting times are
exponential, we may define the current $\tilde{\jmath}_t=\tilde{\jmath}(\C_t)$
associated to the exchange process $\tilde{W}_t$.  The generic formula
\eqref{eq:current} specializes to $\tilde{\jmath}(m)=\frac{1}{\longc}l_m^{-1}
(p_m^+-p_m^-)$. The average $\Espst{\tilde{\jmath}_t}$ of $\tilde{\jmath}_t$ in
the stationary state is thus $\Espst{\tilde{\jmath}_t}= \frac{1}{\longc}\sum_m
l_m^{-1} (p_m^+-p_m^-) \Probst(m)$. The equation for $\Probst(m)$ is
\[
l_m^{-1}\Probst(m)=l_{m+1}^{-1}p_{m+1}^- \Probst(m+1)+l_{m-1}^{-1}p_{m-1}^+
\Probst(m-1),
\] 
which by a simple rearrangement (using $1=p_m^++p_m^-$ on the left-hand side)
can be rewritten:
\[ l_m^{-1}p_m^+\Probst(m)-l_{m+1}^{-1}p_{m+1}^- \Probst(m+1)
=l_{m-1}^{-1}p_{m-1}^+\Probst(m-1)-l_{m}^{-1}p_{m}^- \Probst(m).\] Thus the
quantity $l_m^{-1}p_m^+\Probst(m)-l_{m+1}^{-1}p_{m+1}^- \Probst(m+1)$ is
$m$-independent, i.e. constant along the cycle, and we call it $\Delta^+$. Note
that summing over $m$ one finds that $\Delta^+=\Espst{\tilde{\jmath}_t}$.
Moreover, treating $\Delta^+$ as a parameter, the one term recursion relation
can be solved as before: one iterates until a closed equation is obtained
because $\Probst(0)=\Probst(\longc)$. Then $\Delta^+$ is obtained by the
normalization condition $\sum_m \Probst(m)=1$. One finds without surprise that
the formula for $\Espst{\tilde{\jmath}_t}$ is exactly the formula obtained in
\eqref{eq:afpt} for $1/\Esp{T^+}$, so that
$\Espst{\tilde{\jmath}_t}=1/\Esp{T^+}$ in agreement with the more general result
$\lim_{t \to +\infty} \tilde{W}_t/t = 1/\Esp{T^+}$ obtained above because
$\Esp{\tilde{W}_t}\sim t \Espst{\tilde{\jmath}_t}$ for large $t$.

\subsubsection{Illustration in the degenerate case $n=1$}

Though real cycles have $\longc \geq 3$ vertices, there is nothing that prevents
defining the winding number process for $\longc=1$, so that all indices can be
suppressed. At each jump the winding number grows by $\pm 1$ with probability
$p^{\pm}$. So the corresponding Markov chain is nothing but the simple
asymmetric random walk, and the renewal process is obtained by using a sequence
of independent identically distributed waiting times separating the jumps.
Assume for definiteness that $p^+ \geq p^-=1-p^+$ so that $A=\ln (p^+/p^-)
\geq 0$.  Trivially \[\mathbb{F} =\left(\begin{smallmatrix} 1/(xp^+)& -p^-/p^+ \\
    1 & 0\end{smallmatrix}\right).\] Applying the general formul\ae\ obtained
above,
one recovers very classical results:\\
-- The probability to reach winding number $w$ is $1$ for $w=0,1,2,\cdots$ while
it is $e^{wA}$ for $w=0,-1,-2,\cdots$.\\
-- The generating functions for the number of steps it takes to reach winding
number $w=\pm 1$ are $F^{\pm}(z)$ given by
\[ F^+(z)=\frac{1-\sqrt{1-4z^2p^+p^-}}{2zp^-}, \;
F^-(z)=\frac{1-\sqrt{1-4z^2p^+p^-}}{2zp^+},\] where the square root is the
standard branch ($\sqrt{1}=1$), and the small $z$ expansion
\[\frac{1-\sqrt{1-4z^2p^+p^-}}{2zp^-}=\sum_{k\geq 0}
(p^+)^{k+1}(p^-)^{k}\frac{(2k)!}{k!(k+1)!}z^{2k+1}\] allows to retrieve the
familiar fact that the number of walks that start from $0$ and reach $1$ for the
first time after $2k+1$ steps is given by the Catalan number
$\frac{(2k)!}{k!(k+1)!}$, each of these walks having weight
$(p^+)^{k+1}(p^-)^{k}$.\\
-- If $x(\lambda)$ is the Laplace transform of the distribution of the waiting
time between two jumps, the Laplace transform of the distribution of the first
passage time at $\pm 1$ is  $F^{\pm}(\lambda)$, where
\[ F^+(\lambda)=\frac{1-\sqrt{1-4x(\lambda)^2p^+p^-}}{2x(\lambda)p^-}, \;
F^-(\lambda)=\frac{1-\sqrt{1-4x(\lambda)^2p^+p^-}}{2x(\lambda)p^+},\]
-- The average of the first passage time at $-1$ is infinite, but the average
first passage time at $1$ is
\[ \Esp{T^+}=\frac{l}{p^+-p^-},\] where $l$ is the average time between two jumps, and
$\tilde{W}_t\sim \frac{p^+-p^-}{l}t$ at large $t$.

\section{Conclusions}

In this article, we have proved a new general fluctuation relation dual to the
ones usually exhibited in the literature. 

The usual question is to ask whether there is a relation between the
probabilities that an observable $\tilde{W}_t$ (here, the winding number) takes
a certain value $w$ versus $-w$ at a fixed time $t$: time is fixed,
$\tilde{W}_t$ is random and one compares in some way $p(\tilde{W}_t=w)$ and
$p(\tilde{W}_t=-w)$.  In our approach, we have defined $T^{(w)}$ as the random time
it takes for the observable to reach a certain value $w$ and we compared
the laws of $T^{(w)}$ and $T^{(-w)}$ and derived a corresponding non-equilibrium
fluctuation relation. Technically, the main probabilistic tool is the strong
Markov property. It leads to recursion relations among the quantities of interest,
typically generating functions. Then periodicity along the cycle is used to
solve the recursion relations.

Though this fluctuation relation is established for special systems with the
geometry of a cycle, we have argued that it can be observed in any Markovian
system because the quantity involved in the fluctuation relation is the (cycle)
affinity, which we have shown to be invariant under many probabilistic
contructions, including conditioning but also ``drag and drop'', a new procedure
we have introduced.

One obvious useful generalization would be the study of correlations of random
times between several cycles.
 
\appendix

\section{A short reminder on graph theory} \label{sec:srgt}

In this section, which is meant to be self-contained (see the first chapters of
e.g.
\cite{vanlint-wilson1992,bollobas1998,korte-wygen2008,heineman-pollice-selkov2009}
for much more material), we recall basic definitions, sometimes adapted to our
needs, from elementary graph theory.

A graph $\mathbf{G}$ is a couple $(\mathbf{V},\mathbf{E})$ where $\mathbf{V}$ is
an arbitrary non-empty set and $\mathbf{E}$ is an arbitrary subset of
$\mathbf{V}\times\mathbf{V}$ disjoint from the diagonal $\{(v,v), v\in
\mathbf{V}\}$. Elements of $\mathbf{V}$ are called vertices and elements of
$\mathbf{E}$ are called edges: if $(v,v')\in \mathbf{E}$ (which implies that $v$
and $v'$ are distinct) we say that there is an edge from $v$ to $v'$, or that
$v'$ is adjacent to $v$.  In pictures, the vertices are represented as points,
and the edges as arrows joining vertices. For our purposes, the set of vertices
will always be a finite set.

A non-oriented graph $\mathbf{G}$ is a couple $(\mathbf{V},\mathbf{E})$ where
$\mathbf{V}$ is an arbitrary non-empty set and $\mathbf{E}$ is a subset of the
set, sometimes denoted $S^2(\mathbf{V})$, of pairs of elements of $\mathbf{V}$,
i.e. the set $\big\{\{v,v'\}, v,v' \in \mathbf{V}, v\neq v'\big\}$. In that case,
edges are represented as line segments because in $\{v,v'\}$, $v$ and $v'$ play
symmetric roles.

Graphs such that $\mathbf{E}$ is symmetric, i.e. such that $(v,v')$ is an edge
if and only if $(v',v)$ is an edge, are said to be \textit{micro-reversible}.
This is not standard terminology, but is motivated by statistical mechanics
considerations. Micro-reversible graphs are in one-to-one correspondence with
non-oriented graphs. In pictures, a line segment between to vertices
representing the edge $\{v,v'\}$ corresponds to a pair of arrows, one from $v$
to $v'$ for $(v,v')$ and one from $v'$ to $v$ for $(v',v)$.

A graph $\mathbf{H}$ is a subgraph of a graph $\mathbf{G}$ if the set of
vertices of $\mathbf{H}$ is a subset of the set of vertices of $\mathbf{G}$, and
the set of edges of $\mathbf{H}$ is a subset of the set of edges of $\mathbf{G}$
joining vertices of $\mathbf{H}$.

\begin{figure}[h!]
  \centering
 \includegraphics[width= .9\textwidth]{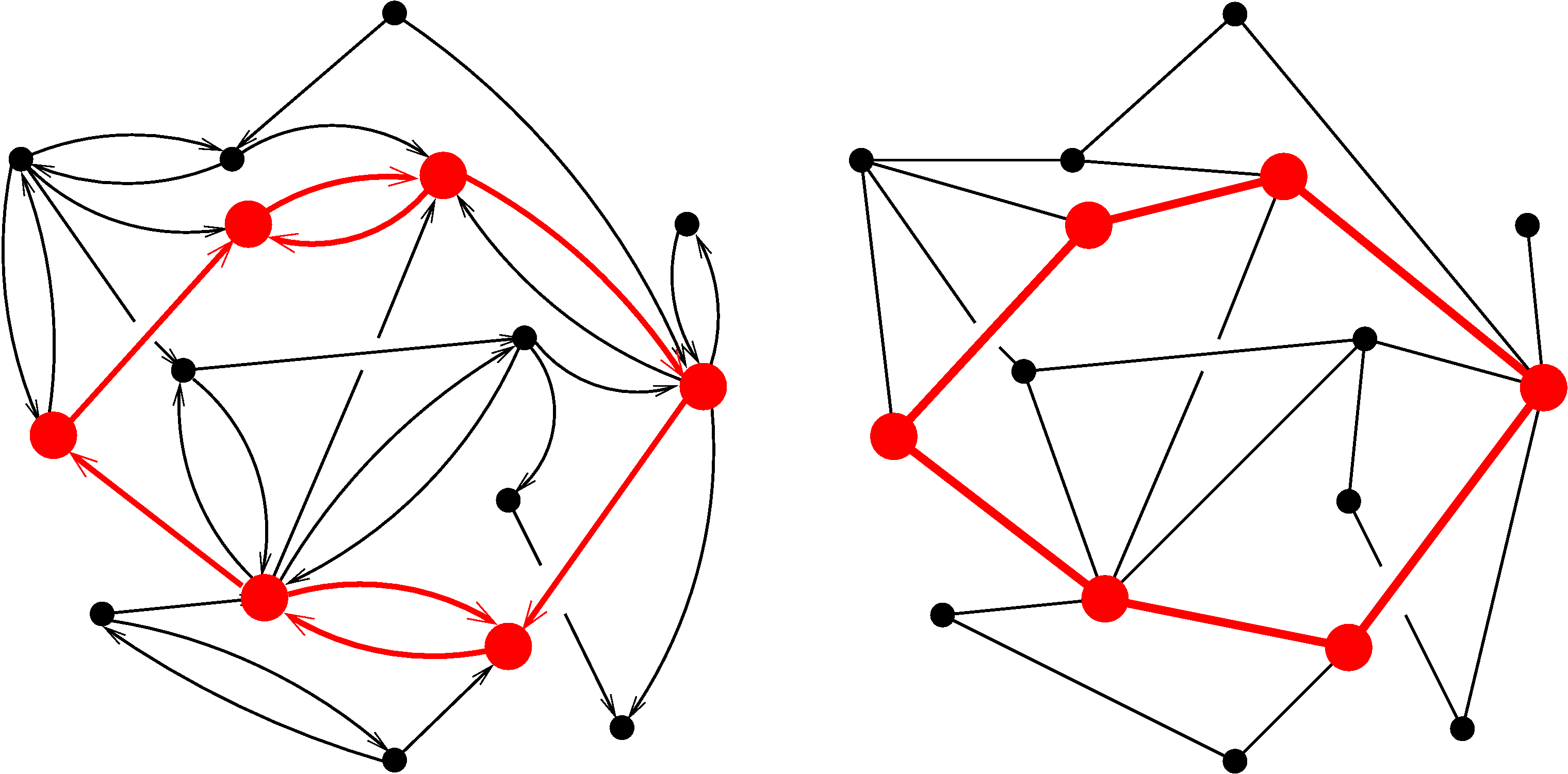}
 \caption{Two examples of subgraphs within a graph. Left: a general graph with a
   subgraph. Right: a non-oriented graph with a subgraph. The subgraph is marked
   with larger vertices and thicker edges. The oriented graph on the left is not
   connected, but the subgraph is.}\label{subgraph}.
\end{figure}

A walk on a graph $\mathbf{G}$ is a sequence $X_0,\cdots,X_N$ of vertices of
$\mathbf{G}$ such that $(X_{i-1},X_{i})$ is an edge of $\mathbf{G}$ for
$i=1,\cdots,N$. The number $N$ is the length of the walk, and single vertices
count as walk of length $0$.

We say that a graph is connected if there is a walk joining any two vertices.
Caution: this is not the standard terminology, one of the reasons being that
micro-reversible graphs $\mathbf{G}$ can be split in a unique way into disjoint
connected components, but this is not true for general graphs.

\section{A short reminder on finite state Markov chains} \label{ShortRemindMC}

A reference dedicated to (finite or countable state space) Markov chains and
processes, mostly self-contained and at an accessible level of sophistication,
is \cite{stroock2005}. Markov chains and processes are also covered in a number
of general textbooks. We have found,
\cite{grimmet-stirzaker2001,koralov-sinai2007,klenke2008,fristedt-gray1997,kallenberg2001},
(at increasing level of complexity and abstraction) well-adapted to our tastes.
But this appendix should be easily readable with a very modest background in
finite probability spaces, and the above references are needed only for those
who want to learn more. The only deep point, namely that the law for infinite
trajectories can be constructed as a limit of laws for finite trajectories (a
special case of Kolmogorov's theorem, see the references above), can be
admitted.

The starting point is a finite set $\mathbf{C}$ of configurations or states, and
a stochastic matrix $\Pdt$ on $\mathbf{C}$. 

A stochastic matrix on $\mathbf{C}$ is a collection $(\C'\vert \Pdt \vert
\C)_{\C',\C \in \mathbf{C}}$ such that $(\C'\vert \Pdt \vert \C)\geq 0$ for
$\C,\C'\in \mathbf{C}$, and $\sum_{\C'} (\C'\vert \Pdt \vert \C)=1$ for $\C \in
\mathbf{C}$. So each column of $\Pdt$ is a probability measure on $\mathbf{C}$.

To these data, we associate a Markov chain on $\mathbf{C}$ as follows: we give
an initial probability distribution $\Pin$ on $\mathbf{C}$ at time $n=0$, and if
at time $n$ the system is in configuration $\C$ then (whatever happened before
$n$) at time $n+1$ the system will be in configuration $\C'$ with probability
$(\C'\vert \Pdt \vert \C)$. One can show (using Kolmogorov's theorem for
instance, see references above) that this defines uniquely a probability measure
on sequences $(\C,\C',\C'',\cdots)$ of elements of $\mathbf{C}$. The Markovian
character is transparent in this description: to know the fate of the system at
time $n+1$, only the knowledge of the system at time $n$ is needed.

It is convenient to have a compact notation and we sometimes write ``
Consider a Markov chain $(\mathbf{C},\Pdt)$ '' or if the initial probability
distribution is important ``Consider a Markov chain $(\mathbf{C},\Pdt,\Pin)$''.
We also sometimes write $\C_n$ ($n=0,1,\cdots$) for the configuration of
the chain at time $n$, i.e. after $n$ steps.

\subsection{Description of trajectories} \label{ssec:tdc}

There are (at least) two natural descriptions of a trajectory:\\
$\bullet$ A first one is: the system is at $\C$ at time $0$, at $\C'$ at time $1$,
at $\C''$ at time $2$, and so on. The corresponding probabilities are easy to
compute. For instance the probability to be at $\C$ at time $0$, at
$\C'$ at time $1$, and at $\C''$ at time $2$ is
\[(\C''\vert \Pdt \vert \C')(\C'\vert \Pdt \vert \C) (\C \vert \Pit) \] where
$(\C \vert \Pit)$ is the initial probability distribution. \\ 
$\bullet$ A second one is:
the system is at $\C$ at time $0$ and stays at $\C$ until time $n$, but jumps
between time $n$ and time $n+1$ to $\C'\neq \C$ and stays at $\C'$ up to time
$n+n'+1$, but jumps between time $n+n'+1$ and time $n+n'+2$ to $\C''\neq \C'$ and so
on. The corresponding probabilities are again easy to compute. For instance the
probability up to the second jump is
\[(\C''\vert \Pdt \vert \C')(\C'\vert \Pdt \vert \C')^{n'} (\C'\vert \Pdt \vert
\C) (\C\vert \Pdt \vert \C)^n (\C \vert \Pit). \]  

The second description is related to the following construction.  Define a new
stochastic matrix $\Ppdt$ by the formul\ae : \\
-- If $(\C \vert \Pdt \vert \C)=1$ then $(\C' \vert \Ppdt \vert \C)=(\C' \vert
\Pdt \vert \C)$ for every $\C'$ (i.e. $(\C \vert \Ppdt \vert \C)=1 $ and $(\C'
\vert \Ppdt \vert \C)=0$ for $\C'\neq \C$).\\
-- If $(\C \vert \Pdt \vert \C)<1$ then $(\C \vert \Ppdt \vert \C)=0$ and $ (\C'
\vert \Ppdt \vert \C)=\frac{(\C'
  \vert \Pdt \vert \C)}{1-(\C \vert \Pdt \vert \C)}$ for $\C'\neq \C$. \\
Also let $\K$ be the diagonal matrix whose entries are those of the diagonal of
$\Pdt$, i.e.  $(\C'\vert \K \vert \C)\equiv \delta_{\C,\C'} (\C\vert \Pdt \vert
\C)$.  Then \[\Ppdt (\Id-\K)+\K=\Pdt. \]

Say that a random variable $N$ obeys a geometric distribution of parameter $k\in
[0,1]$ if $p(N\geq n)= k^n$ for $n=0,1,\cdots$, i.e. $p(N=n)= k^n(1-k)$. Then
the probabilistic interpretation of the second description is the following.
Starting at $\C$ at time $0$ if $(\C\vert \K \vert \C)=1$ stay there forever,
otherwise stay at $\C$ during a geometric time $N$ with parameter $(\C\vert \K
\vert \C)$ and between $N$ and $N+1$ jump to $\C'$ with probability $(\C' \vert
\Ppdt \vert \C)$; if $(\C'\vert \K \vert \C')=1$ stay there forever, otherwise
stay at $\C'$ during a geometric time $N'$ with parameter $(\C'\vert \K \vert
\C')$ and between $N+N'+1$ and $N+N'+2$ jump to $\C''$ with probability $(\C''
\vert \Ppdt \vert \C')$; and so on.

Note that certain (infinite) trajectories for the Markov chain associated to
$\Pdt$ may lead to a finite number of jumps. However, if $(\C\vert \K \vert
\C)=1$ then also $(\C\vert \Ppdt \vert \C)=1$ so staying forever at $\C$ is also
what is predicted by $\Ppdt$, and if this is done, any infinite trajectory for
the Markov chain associated to $\Pdt$ leads to an infinite trajectory for the
Markov chain associated to $\Ppdt$.

\subsection{The graph associated to a Markov chain} \label{ssec:gaMc}
 
To any stochastic matrix $\Pdt$ we can associate a graph. The vertices
are the configurations in $\mathbf{C}$, and (oriented) edges are indexed by the
possible transitions: there is an oriented edge from $\C$ to $\C'$ if and only
if $\C \neq \C'$ and $(\C'\vert \Pdt \vert \C) \neq 0$. In that case, we talk of
the edge $(\C,\C')$ and we say that $\C'$ is adjacent to $\C$ (not a symmetric
relation in general). We shall denote the corresponding graph by $\mathbf{G}$:
as recalled above, $\mathbf{G}$ is a couple of sets: the set of vertices,
$\mathbf{C}$, and the set of edges, a subset of $\mathbf{C} \times \mathbf{C}$
disjoint from the diagonal.

Recall that a graph $\mathbf{G}$ is said to be connected if one can go from any
vertex to any other by a sequence of adjacent vertices. In that case, general
theorems state that:\\ 
-- All configurations are recurrent with probability $1$, i.e. for almost every
trajectory, every configuration appears infinitely many times.\\
-- There is a single probability measure $\Probst$ on $\mathbf{C}$, called the
stationary measure, such that $\Pdt \Probst =\Probst$ i.e. $\sum_{\C} (\C'\vert
\Pdt \vert \C)\Probst(\C)=\Probst(\C')$ for each $\C'$.  Moreover,
$\Probst(\C)>0$ for each $\C\in \mathbf{C}$. 

However, due to some possible
arithmetic symmetries, is is not always true that any initial probability
distribution on $\mathbf{C}$ converges at large times to $\Probst$.

Motivated by the graphical interpretation, we shall repeatedly use the image
that a trajectory is a random walker on $\mathbf{C}$ jumping from time to time
along edges from one vertex to another. Note that $\Pdt$ and the associated
$\Ppdt$ correspond to the same graph.

We shall often make the assumption of micro-reversibility, i.e. $(\C'\vert \Pdt
\vert \C)$ and $(\C\vert \Pdt \vert \C')$ are simultaneously either $=0$ or
$\neq 0$. Then, the above (oriented) graph carries the same information as a
non-oriented one, for which we keep the same name, and we shall say that
$\{\C,\C'\}$ (unordered) is an edge of $\mathbf{G}$.

\section{A short reminder on  finite state Markov processes}
\label{ShortRemindMP}

This appendix should be easily readable by anyone with minimal familiarity with
the master equation approach.  Our discussion closely parallels the description
of Markov chains. The references given at the beginning of Appendix
\ref{ShortRemindMC} explain what is stated here and much more.

The starting point is a finite set $\mathbf{C}$ of configurations or states, and
a transition matrix $\Trans$ with vanishing diagonal matrix elements and whose
non-diagonal matrix elements $(\C'\vert \Trans \vert \C)\geq 0$ , $\C,\C'\in
\mathbf{C}$, $\C \neq \C'$, describe transition rates from $\C$ to $\C'$. To
$\Trans$ is associated a Markov matrix $\M$ defined by $(\C'\vert \M \vert
\C)\equiv (\C'\vert \Trans \vert \C)$ for $\C,\C'\in \mathbf{C}$, $\C \neq \C'$,
and $(\C\vert \M \vert \C)\equiv -\sum_{\C' \neq \C} (\C'\vert \Trans \vert
\C)$ for $\C \in \mathbf{C}$. We let $\D$ denote the (diagonal) matrix such that
$\M=\Trans-\D$.  

To these data, we associate a Markov process on $\mathbf{C}$ as follows: we give
an initial probability distribution $\Pin$ on $\mathbf{C}$ at time $t=0$, and if
at time $t$ the system is in configuration $\C$ then (whatever happened before
$t$) at time $t+\Delta t$ the system is still in $\C$ with probability
$1+(\C\vert \M \vert \C)\Delta t + o(\Delta t)$ i.e.  $1-(\C\vert \D \vert
\C)\Delta t + o(\Delta t) $, and is in configuration $\C' \neq \C$ with
probability $(\C'\vert \M \vert \C)\Delta t + o(\Delta t)$ i.e.  $(\C'\vert
\Trans \vert \C)\Delta t + o(\Delta t)$. Taking an infinitesimal $dt$ instead of
a finite but small $\Delta t$, we see that $\M$ appears as the infinitesimal
generator of the Markov process. The Markovian character is transparent in this
description: to know the fate of the system at time $t+dt$, only the knowledge
of the system at time $t$ is needed. The mathematical construction of the law of
a Markov process from these data is only slightly more complicated than for
Markov chains.

\subsection{Relation with the master equation} \label{ssec:rwme}

We write $\C_t$ ($t\in [0,+\infty[$) for the configuration of the process at
time $t$. Thus a trajectory is a (random) function from $[0,+\infty[$ to
$\mathbf{C}$.  The mathematical construction of the Markov process amounts to
the construction of a probability measure, which we denote by $\Prob$ in this
article, consistent with the heuristic description of transitions given above,
on an appropriate set of functions from $[0,+\infty[$ to $\mathbf{C}$. One shows
that it is possible to concentrate on functions which are right-continuous with
left limits: if there is a jump at time $t$, $\C_t$ is the position after the
jump so $\C_t\neq \C_{t^-}$ if and only if there is a jump at time $t$. Moreover
one can concentrate on functions that have only a finite number of jumps during
any bounded time interval. So if $\Q$ is any matrix indexed by $\mathbf{C}$ and
with vanishing diagonal elements, ($\Trans$ is a typical example), the sum
$\sum_{t\in ]0,T]} (\C_t \vert \Q \vert \C_{t^-})$, to which only jump times
contribute, is well-defined. This simple construction plays an important role in
the construction of cumulative processes in section \ref{ssec:cp}.

It is convenient to have a compact notation and we sometimes write
``Consider a Markov process $(\mathbf{C},\Trans)$'' or if the initial
probability distribution is important ``Consider a Markov process
$(\mathbf{C},\Trans,\Pin)$''.

The probability $\Prob(\C;t)$ to be in configuration $\C$ at time $t$, a
shorthand for $\Prob(\C_t=\C)$, is obtained by solving the so-called master
equation
\[\frac{d}{dt}\Prob(\C;t)=\sum_{\C'} (\C\vert \M \vert \C')\Prob(\C';t).\]
with initial condition $\Prob(\C;0)$.
This can also be written as 
\[\frac{d}{dt}\Prob(\C;t)=\sum_{\C'\not=\C}(\C\vert \Trans\vert \C')
\Prob(\C';t)-\sum_{\C' \not=\C}(\C'\vert \Trans\vert \C) \Prob(\C;t) \equiv
\sum_{\C'} \dot{\Prob}_{\C;\C'}(t)\] which has a simple interpretation:
$\Prob(\C;t)$ varies with time because of positive contribution due to jumps to
$\C$ and of negative contributions due to jumps from $\C$, and for the
infinitesimal time balance this leads to the above formula. By construction
$\dot{\Prob}_{\C;\C'}(t)\equiv (\C\vert \Trans\vert \C') \Prob(\C';t)-(\C'\vert
\Trans\vert \C) \Prob(\C;t)$, which contains the contributions of transition from
$\C'$ or to $\C'$ to the variation of $\Prob(\C,t)$,  is anti-symmetric.

\subsection{The graph associated to a Markov process} \label{ssec:gaMp}

To any Markov matrix $\M$ we can associate a graph. The vertices are the
configurations in $\mathbf{C}$, and oriented edges are indexed by the possible
transitions: there is an oriented edge from $\C$ to $\C'$ if and only if
$(\C'\vert \Trans \vert \C) \neq 0$. In that case, we talk of the edge
$(\C,\C')$ and we say that $\C'$ is adjacent to $\C$ (not a symmetric relation
in general). We shall denote the corresponding graph by $\mathbf{G}$. As usual
$\mathbf{G}$ is a couple of sets: the set of vertices, $\mathbf{C}$, and the set
of edges, a subset of $\mathbf{C} \times \mathbf{C}$.

Remember the graph $\mathbf{G}$ is said to be connected if one can go from any
vertex to any other by a sequence of adjacent vertices. In that case, a general
theorem guaranties that there is a single probability measure $\Probst$ on
$\mathbf{C}$, called the stationary measure, such that $\M \Probst =0$ i.e.
$\sum_{\C'} (\C\vert \M \vert \C')\Probst(\C')=0$ for each $\C$.  Moreover,
$\Probst(\C)>0$ for each $\C$ and whatever $\Prob(\C;0)$, $\lim_{t\to \infty}
\Prob(\C;t)=\Probst(\C)$. Hence the stationary measure is unique, charges every
vertex, and is the infinite time limit of every initial probability
distribution.

Motivated by the graphical interpretation, we shall repeatedly use the image
that a trajectory is a random walker on $\mathbf{C}$ jumping from time to time
along edges from one vertex of  $\mathbf{G}$ to an adjacent vertex.

We shall often make the assumption of micro-reversibility, i.e. $(\C'\vert
\Trans \vert \C)$ and $(\C\vert \Trans \vert \C')$ are simultaneously either
$=0$ or $\neq 0$. Then, the above (oriented) graph carries the same information as
a non-oriented one, for which we keep the same name. 

\subsection{Description of trajectories} \label{ssec:tdp}

There is a more explicit description of trajectories which is useful for our
purpose (and for numerical simulations of trajectories as well). It is the
continuous time analog of the second description of trajectories for Markov
chains, and it goes as follows. Recall that $\mathbb{D}$ is the diagonal matrix
whose entries are those of the diagonal of $\mathbb{M}$.
\\
-- The configuration at $t=0$ is sampled according to the initial probability
distribution $\Pin$ on $\mathbf{C}$. Say the configuration at $t=0$ is $\C$.
\\
-- If $(\C\vert \D \vert \C)=0$ stay in $\C$ forever, otherwise wait an
exponential time $T$ with parameter $(\C\vert \D \vert \C)$ (i.e.
$p(T>t)=e^{-(\C\vert \D \vert \C)t}$) and at $t=T$ jump to the configuration
$\C' \neq \C$ with probability $\frac{(\C'\vert \Trans \vert \C)}{(\C\vert \D
  \vert \C)}$.
\\
-- If $(\C'\vert \D \vert \C')=0$ stay in $\C'$ forever, otherwise wait an
exponential time $T'$ with parameter $(\C'\vert \D \vert \C')$ and at $t=T+T'$
jump to configuration $\C''\neq \C'$ with probability $\frac{(\C''\vert \Trans
  \vert \C')}{(\C'\vert \D \vert \C')}$.
\\
-- ...

To the Markov matrix $\M$ we can associate a stochastic matrix $\Pdt$
(i.e. the generator for a discrete time Markov chain) as follows : \\
-- If $(\C\vert \M \vert \C)=0$ then $(\C\vert \Pdt \vert \C)=1$ and $(\C'\vert
\Pdt \vert \C)=0$ for $\C' \neq \C$. \\
-- Else, $(\C'\vert \Pdt \vert \C)=\frac{(\C'\vert \Trans \vert \C)}{(\C\vert
  \D \vert \C)}$; in particular  $(\C\vert \Pdt \vert \C)=0$.\\
Note that \[\M=(\Pdt-\Id)\D\] and that $\Pdt$ and $\M$ define the same graph.
One important consequence of the above description is that the sequence
$(\C,\C',\C'',\cdots)$ is a sample of the Markov chain associated to $\Pdt$
(with the same innocent trick as in Appendix \ref{ShortRemindMC} in force: if
$(\C,\C',\C'',\cdots)$ is a finite sequence, then one turns it into an infinite
one by repeating its last term over and over).

\subsection{Perturbative expansion} \label{ssec:pe}

For the reader unfamiliar with the above trajectory description of the Markov
process, we can offer a poor man's heuristic version, reminiscent of Feynman's
sum over histories, and which is again  a continuous time analog of the  Markov
chain case. Let $\Uev(t)=e^{\M t}$ be the evolution operator, i.e. the
matrix solution of
\[\frac{d}{dt}\Uev= \M\Uev\]
with initial condition $\Uev(0)=\Id$. Using the identity $\M=(\Pdt-\Id)\D$, it
is easy to see that these two equations can be rephrased as a single integral
equation:
\[\Uev(t)=e^{-\D t} +\int_0^t ds \, e^{-\D(t-s)}\Pdt \D \Uev(s).\]
We can iterate this equation, i.e. inject $U(s)=e^{-\D s} +\int_0^s dr \,
e^{-\D(s-r)}\Pdt \D \Uev(r)$ in the right-hand side, and go on. This gives a
series expansion 
\be \label{eq:evpos} \Uev(t)=e^{-\D t} +\int_0^t ds \,
e^{-\D(t-s)}\Pdt \D e^{-\D s} + \int_0^t ds \, \int_0^s dr \, e^{-\D(t-s)}\Pdt
\D e^{-\D(s-r)}\Pdt \D e^{-\D r}+\cdots.\ee 
The zeroth order term (where $\Pdt$ does not appear) describes trajectories that
make no jump in $[0,t]$, the first order term (a single integral where
$\Pdt$ appears once) ''sums'' over trajectories that make one jump in $[0,t]$,
the jump time being $s$, the second term (a double integral where $\Pdt$ appears
twice) ''sums'' over trajectories that make two jumps in $[0,t]$, the jump times
being $r$ and then $s$, and so on. This is exactly the prediction of the
trajectory description in terms of jumps governed by $\Pdt$ and exponential
waiting times described by $\D$. Note that, in agreement with dimensional
analysis, there is one $\D$ for each time integration variable: this is because
if the random variable $T$ is such that $p(T>t)=e^{-\lambda t}$ then the density
of $T$ is $ dt\, \lambda e^{-\lambda t}$.

This representation gives an easy proof of an important property of
$\Prob(\C;t)$ when $\mathbf{G}$ is connected: whatever $\Prob(\C;0)$,
$\Prob(\C;t)>0$ for every $\C$ and $t>0$, and in particular $\Probst(\C)>0$ for
every $\C$. Indeed, each term on the right-hand side of \eqref{eq:evpos} has
non-negative matrix elements, and if one can go from $\C$ to $\C'$ via $n$
jumps, the $n^{\text{th}}$ order term (the one with $n$ occurrences of $\Pdt$)
has a strictly positive matrix element between $\C$ and $\C'$. So when
$\mathbf{G}$ is connected, for every $t >0$ every matrix element of $\Uev(t)$ is
$>0$.

To conclude this rapid overview, note that for any $\varepsilon >0$
$\Uev(\varepsilon)$ is a stochastic matrix, and that the sequence
$\C_{n\varepsilon}$, $n=0,1,\cdots$ is a sample of the Markov chain
$(\mathbf{C},\Uev(\varepsilon))$. In this way, many properties of Markov
processes can be (at least heuristically) proven by proving an analog for Markov
chains and letting $\varepsilon \to 0$.

\end{document}